\documentclass[12pt]{article}
\usepackage{latexsym,graphicx,color,times}
\usepackage{colordvi}
\usepackage[portrait,margin=1.0in]{geometry}
\usepackage{amsmath}
\def\pd{\partial}

\def\rfig#1{Fig.\ref{fig:#1}}
\def\rsec#1{Sec.\ref{sec:#1}}
\def\rtab#1{Table \ref{tab:#1}}
\long\def\comment#1{}

\newcommand{\be}{\begin{equation}}
\newcommand{\ee}{\end{equation}}

\def\Rdd{\Black}

\def\req#1{(\ref{eq:#1})}
\def\rfig#1{Fig.\ref{fig:#1}}

\def\macname{hank_strauss}

\def\figdird3d{/Users/{\macname}/Documents/progs/m3d/d3d-lock}

\def\figdirw{/Users/{\macname}/Documents/papers/disruption/rwtm/gnuplot}

\def\figdirm{/Users/{\macname}/Documents/progs/m3d/mst}
\def\figdirp5{/Users/{\macname}/Documents/papers/disruption/rwtm/deltap/paper5}

\def\figdirnstx{/Users/{\macname}/Documents/progs/m3d/nstx}

\def\figdirkstr{/Users/{\macname}/Documents/papers/disruption/rwtm/kstar}
\def\figdirliq{/Users/{\macname}/Documents/papers/disruption/rwtm/liq95}

\def\figdir{./pix}

\begin{document}
\begin{center}
{\Large\bf Eliminating  Tokamak Disruptions with Feedback} \\
{\large H. R. Strauss} \\
{HRS Fusion, West Orange NJ 05072 USA} \\
{email: hank@hrsfusion.com} \\
\end{center}

\long\def\comment#1{}
\def\macname{hank_strauss}
\def\figdirkst{/Users/{\macname}/Documents/progs/m3d/kstar}
\def\figdirw{/Users/{\macname}/Documents/papers/disruption/rwtm/gnuplot}
\def\figdirm{/Users/{\macname}/Documents/progs/m3d/mst}
\def\figdirnstx{/Users/{\macname}/Documents/progs/m3d/nstx}
\def\figdirkstr{/Users/{\macname}/Documents/papers/disruption/rwtm/kstar}
\def\figdirliq{/Users/{\macname}/Documents/papers/disruption/rwtm/liq95}
\def\figdir{.}



\section*{\large{\bf Abstract}}
{\small 
Tokamak disruptions caused by tearing modes can be prevented with magnetic feedback.
A database of DIII-D locked mode disruptions provides criteria for major disruptions. 
The  main criterion is that the
 $q = 2$ rational surface must be
sufficiently close to the resistive wall  surrounding the plasma to interact with it.
This is shown in simulations of a sequence of model equilibria derived from 
the database. 
If the wall 
is ideally conducting, major  disruptions do not occur.
Magnetic feedback  emulates a conducting wall, and  can prevent
major disruptions.  
Feedback coils must be
designed appropriately, otherwise they are not effective in
disruption prevention. Simulations of a model of MST also show 
the $q = 2$ criterion for major disruptions, as well as feedback stabilization.
Feedback offers an attractive alternative to disruption mitigation by impurity injection.
}

\section{\large{\bf Introduction}} \label{sec:intro}
Disruptions have been considered to be a major obstacle to tokamak fusion.
The resistive wall surrounding the plasma has an important role in causing disruptions
\cite{jet21,iter21,d3d22,mst23, finn95,betti}.  
Resistive wall tearing modes (RWTM) 
are tearing modes which interact with the resistive wall.  
The term resistive wall mode (RWM) is reserved for ideal external kink modes
which interact with a resistive wall, and are not treated in this paper.
Feedback can suppress the disruptive effect of the resistive wall.

The main criterion \cite{sweeney2017}  for locked mode disruptions in DIII-D 
 is that $\rho_{q2}$ exceeds a threshold, 
   Here $\rho_{q2}$ is the $q = 2$ radius 
    normalized to wall radius.
It can be understood as the distance required for a tearing mode to 
interact with the resistive wall. 
     
A second criterion is that the internal inductance $ l_i / q_{95}$  exceeds a critical value,
which is correlated with the $\rho_{q2}$ criterion.
This characterizes  equilibria with $q_0 \approx 1,$ and
small toroidal current density for $\rho > \rho_{q2}.$

It is generally assumed that disruptions are caused by tearing modes.
Typically nonlinear tearing modes have been simulated with an ideally conducting wall
surrounding the plasma \cite{white80,waddell}, with a highly unstable initial state.
Such a state does not arise spontaneously in experiments, because
the plasma evolves on a slow time scale, which allows it to relax to a more
stable state. 
It was not generally appreciated that a resistive wall is a very important
factor in causing disruptions.
In this study, derived from data,
the initial states can be  tearing unstable with an ideal wall, but are not 
sufficiently unstable to cause a major disruption.

Nonlinear simulations demonstrate that with a resistive wall,
   RWTMs can  
can cause a complete thermal quench. 
This is because they produce a significant magnetic perturbation at the
plasma edge. This allows heat to be transported across the edge by parallel thermal
conduction. 
In contrast, tearing modes with an ideal
wall boundary condition produce small edge magnetic perturbations, 
which inhibits parallel thermal conduction, permitting only  a minor disruption.

Feedback can emulate an  ideal wall and  prevent major disruptions.
Simulations indicate that the feedback coils must be designed appropriately to
align with the RWTM mode structure at the resistive wall. 

Feedback stabilization was employed successfully \cite{hanson,piovesan}
 in DIII-D and RFX to suppress 
resistive wall modes (RWM), 
with $\rho_{q2} \approx 1.$  Here it is shown that the range of possible disruption
prevention is much wider, extending to all  $\rho_{q2} < 1.$ Above the critical
$\rho_{q2},$ feedback can suppress major disruptions, while below the
$\rho_{q2}$ threshold,  major disruptions do not occur, even though tearing modes can be
unstable.  Feedback in RFX with $\rho_{q2} < 1$ was attempted \cite{zanca} 
but encountered density limit problems.  
Here simulations of DIII-D and MST demonstrate feedback stabilization for
$\rho_{q2} < 1.$

This paper extends previous work \cite{nstx25,nf-iaea24} in 
several ways. Simulations representing the DIII-D locked mode disruption database
 are performed. It it demonstrated that  
major disruptions do not occur with an ideal wall. With a resistive wall, major
disruptions occur when the $\rho_{q2}, l_i / q_{95}$  criteria are satisfied.  Feedback
stabilization is studied with a more realistic model, based on DIII-D I-coils. 
It is shown that it is possible to emulate an ideal wall with feedback.
Simulations using model MST equilibria compare the disruption criteria with DIII-D,
and verify feedback stabilization with coils similar to I-coils.

Feedback stabilization is preferable to the presently preferred method for 
disruption mitigation, which is to inject high Z impurities in order to radiate
energy.  This method causes a disruption in itself \cite{izzo,nardon,spi,spi2}. 
A serious side effect is
excitation of runaway electrons \cite{reux,rosenbluth}, which can severely damage the wall
surrounding the plasma. It seems sensible to avoid  this mitigation where possible.

The paper is organized as follows. \rsec{database} discusses the DIII-D database and
gives a simple intuitive model of the $\rho_{q2}$ criterion.
\rsec{d3dsim} presents a sequence of DIII-D equilibrium simulations with different
$\rho_{q2}.$  It is verified that they satisfy the two criteria in the database when
a resistive wall boundary condition is employed. With an ideal wall, there are no
major disruptions.
\rsec{feedback} describes the feedback model.
\rsec{d3d} applies feedback to disruptive cases in \rsec{d3dsim}. 
Good results are obtained with I-coil feedback  symmetric about the midplane, 
or approximately aligned with the perturbed  magnetic field.
Stabilization is not obtained with I-coil feedback antisymmetric about the midplane.
\rsec{mst} deals with feedback simulations of an MST model.
\rsec{current} discusses the inductive criterion in DIII-D and MST.
The results are summarized in \rsec{summary}.
  
\section{\Rdd {\bf  DIII-D database and  disruption  criteria}} \label{sec:database}
\begin{figure}[h]
 \begin{center}
\includegraphics[height=5.0cm] {\figdir/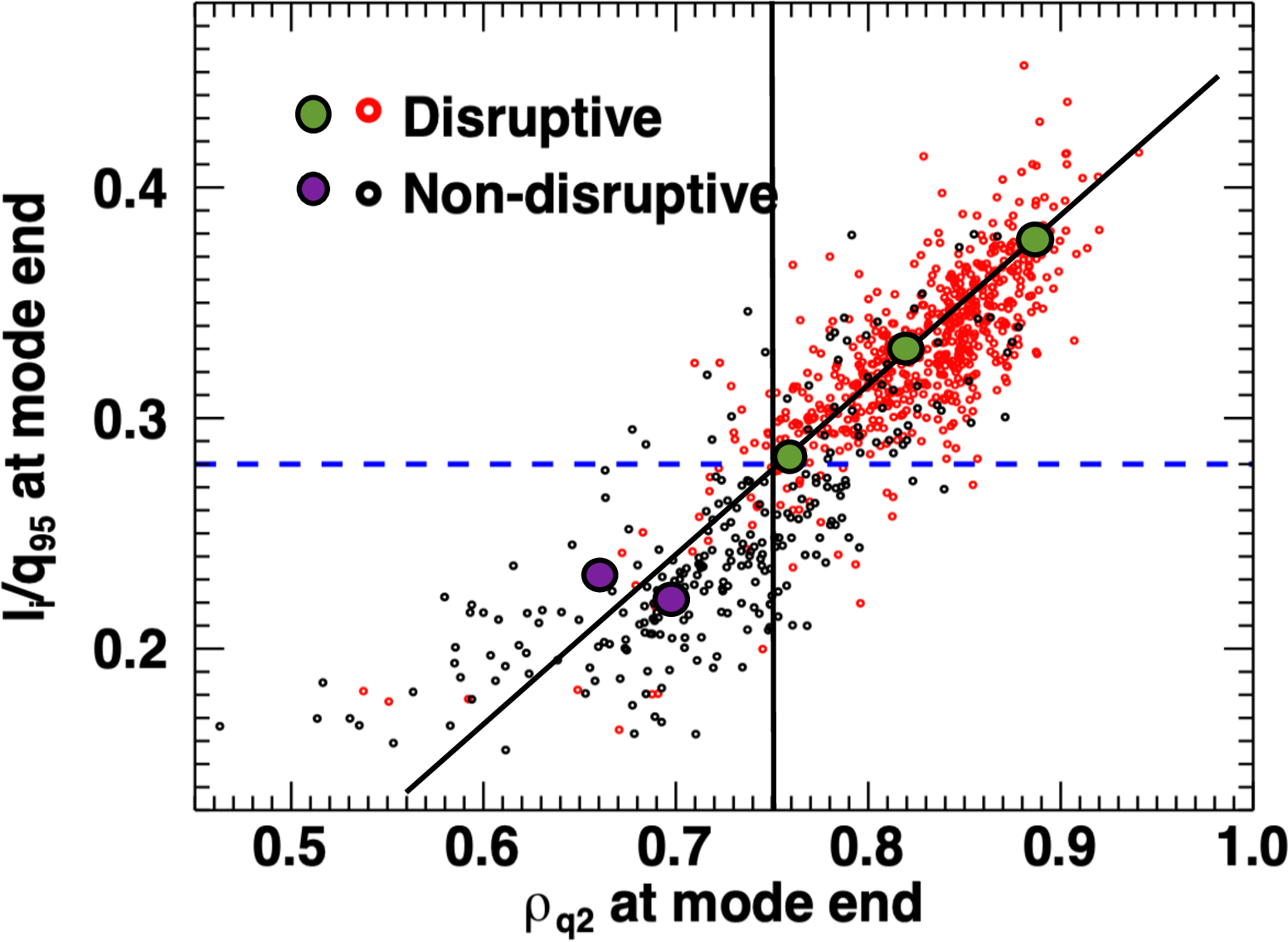}
\end{center}
\caption{\it
  Disruptivity in a DIII-D locked mode disruption database,
adapted from \cite{sweeney2017} with IAEA permission.} \label{fig:sweeney}
\end{figure}

Sheared rotation stabilizes tearing modes. When the rotation locks, tearing modes can become
unstable. 
Disruption precursors cause edge cooling, which causes the current to contract. This
can cause two main disruption criteria to be satisfied.
This is shown in a database \cite{sweeney2017} of DIII-D locked mode disruptions.
\rfig{sweeney} shows disruptivity in the $(\rho_{q2},l_i/q_{95})$ plane. The
horizontal coordinate $\rho_{q2}$ is the minor radius of the $q = 2$ rational
surface, normalized to the plasma radius $a$.
The  disruption criterion  is $\rho_{q2} \ge 0.75$. 
The vertical coordinate is the internal inductance divided 
by $q_{95}.$
The expression ``at mode end" means $100 ms$ before the end of the shot.
The sloping line is a fit \cite{sweeney2017} to the data, 
\be l_i / q_{95} = (2/3) \rho_{q2} - .22 \label{eq:fit} \ee
The $\rho_{q2}$ and $l_i / q_{95}$ criteria are correlated.
The filled circles are simulations presented in \rsec{d3dsim}. The green  circles
have major disruptions if the wall is resistive. The purple circles do not disrupt.

The $\rho_{q2}$ disruption criterion indicates that the modes are RWTMs. 
The  modes are tearing, because $\rho_{q2} < 1,$ and RWTMs, because $\rho_{q2}
> 0.75,$ allowing wall interaction. 
The criterion is the condition for a lobe of the  mode to reach the wall, 
so the mode ``knows" the wall boundary condition.
\begin{figure}[h] 
\begin{center}
\includegraphics[height=4.2cm]{\figdir/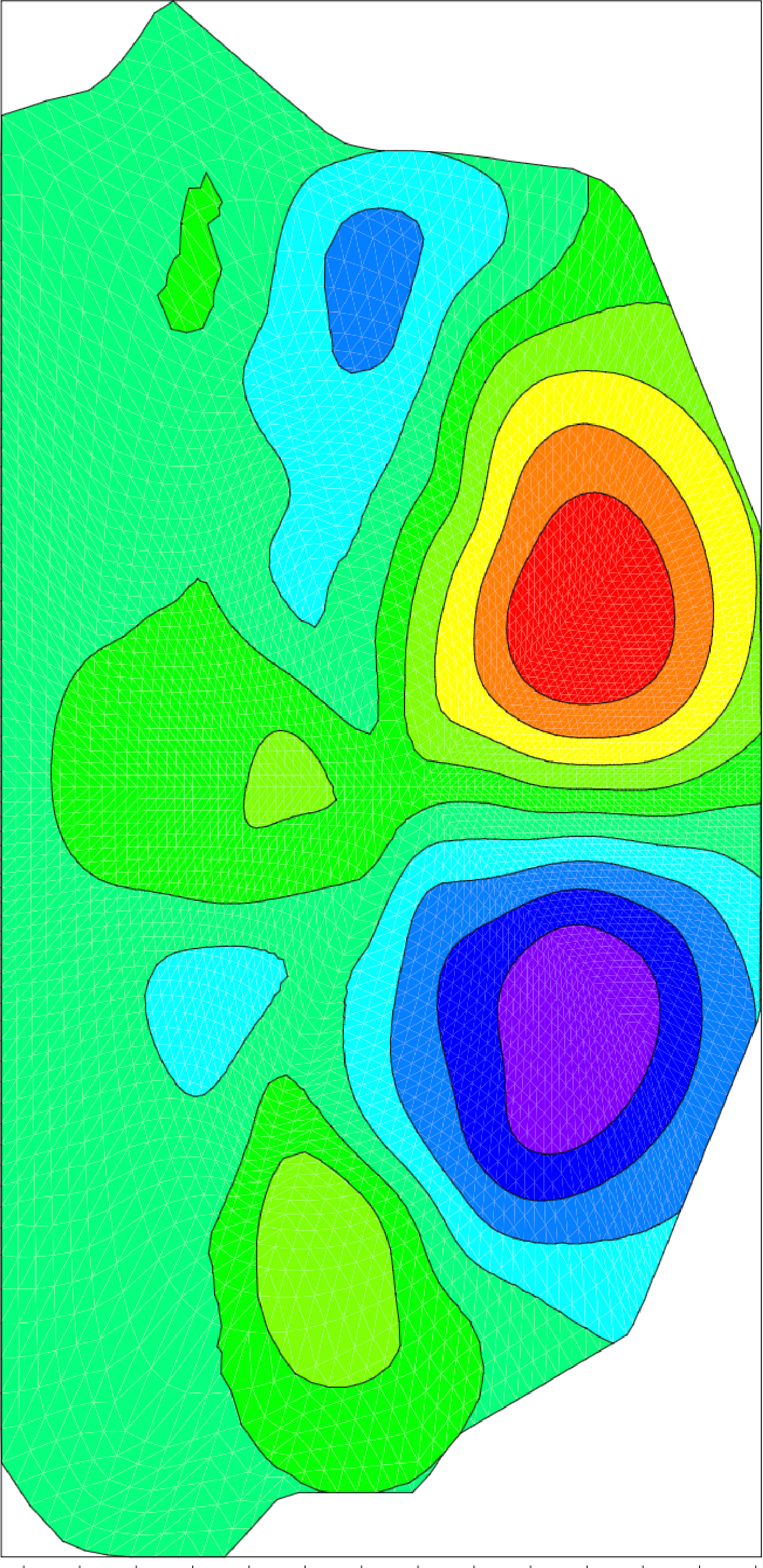}(a)
\includegraphics[height=2.8cm]{\figdir/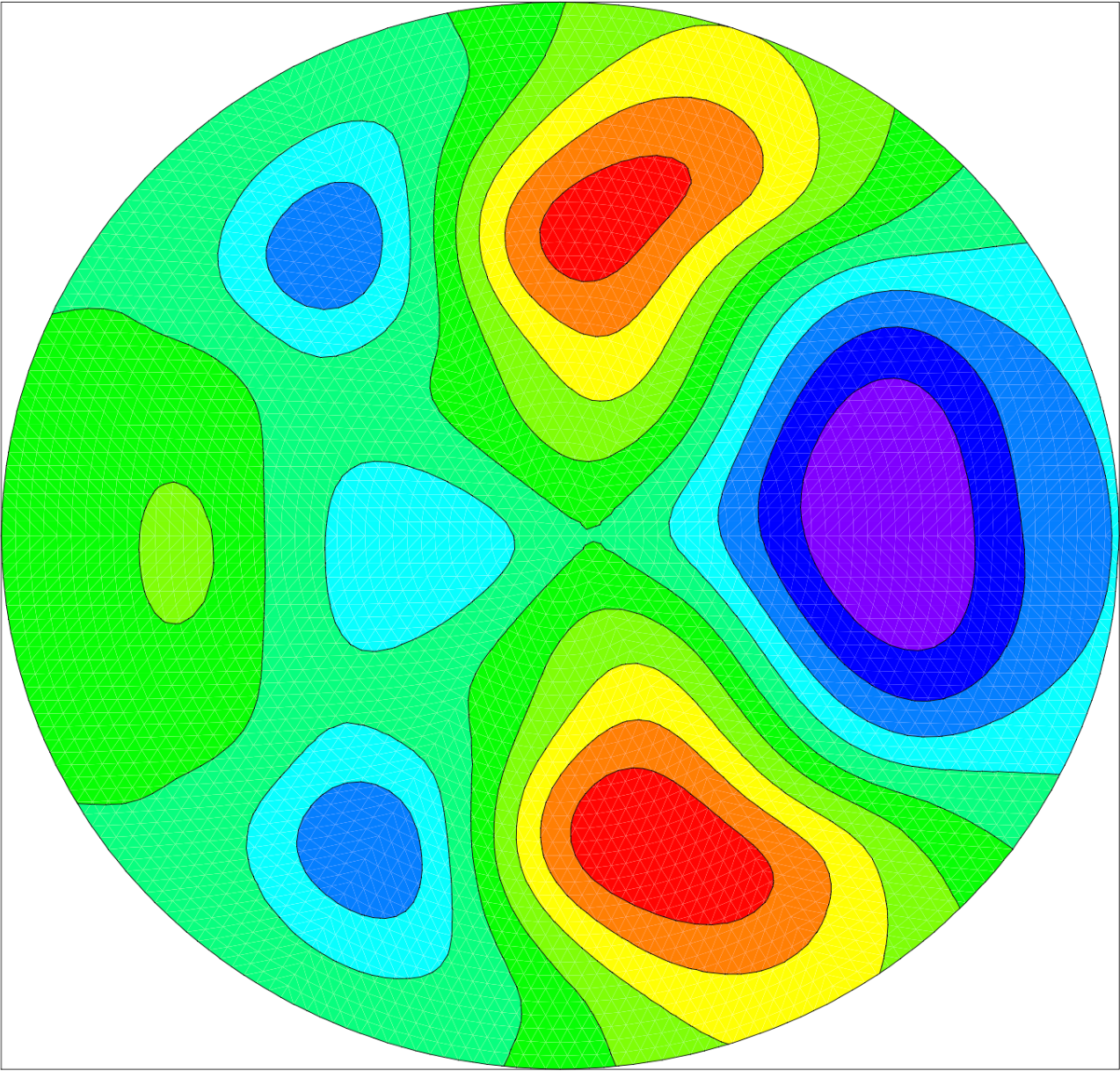}(b)
\includegraphics[height=4.2cm]{\figdir/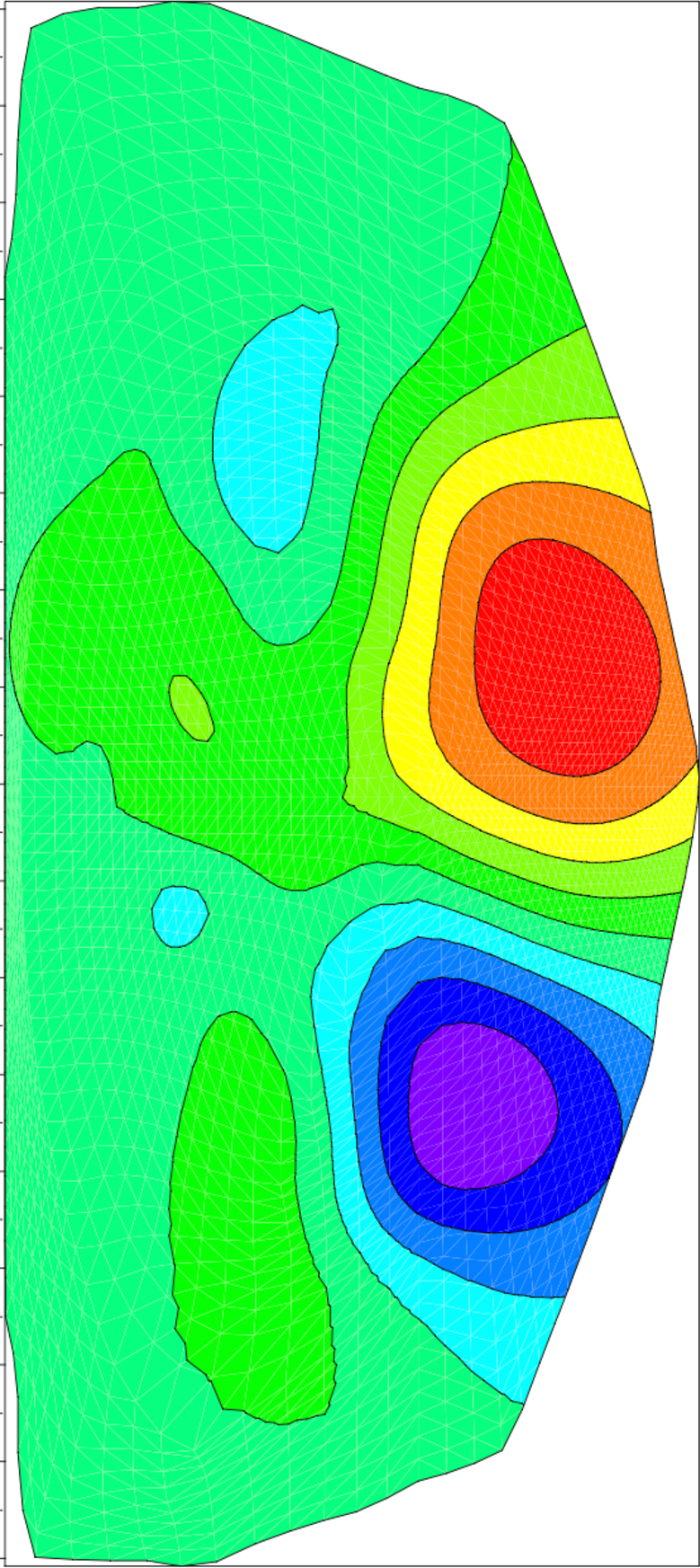}(c)
\includegraphics[height=4.2cm]{\figdir/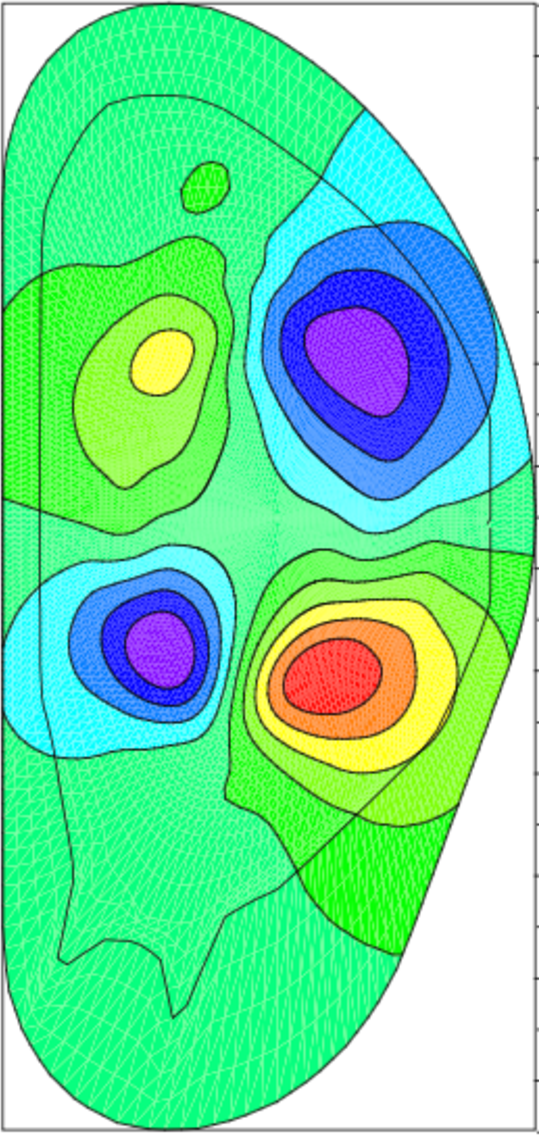}(d)
\end{center}
\caption{\it
(a) Perturbed $\psi$ in DIII-D simulations in \rfig{simbase}(a), $\rho_{q2} = 0.82,$ time
$t = 1820 \tau_A, $ in poloidal plane $\phi = 0.$
(b) Perturbed $\psi$ in MST model  in  \rsec{mst}, reproduced from \cite{model} with AIP permission; (c) NSTX 
reproduced from \cite{nstx25} with IAEA permission, (d) 
ITER \cite{iter21}. }
\label{fig:lobe}
\end{figure}
\rfig{lobe} gives examples of simulations of RWTMs in several experiments: 
(a) DIII-D \cite{d3d22}, (b) Madison Symmetric Torus (MST) \cite{mst23,model}, 
(c) NSTX \cite{sabbagh2010,nstx25}, and (d) ITER \cite{iter21}.
Shown is the perturbed magnetic flux $\psi$ in a poloidal plane. 
The normalized distance to the wall is $\rho_w = \rho_{q2} + 1/ (k_\perp a),$
where the wavenumber is the normalized lobe radius $k_\perp a \approx m / \rho_{q2},$
and $\rho_w$ is the normalized wall radius.
For mode number $m = 2$ 
\be  \rho_{q2} \approx \frac{\rho_w}{1 + 1/2} \approx  0.8 \label{eq:rhoq2} \ee
taking  $\rho_w \approx 1.2$ in examples (a) - (d).
More accurately \cite{nstx25, nf-iaea24, model},
the onset condition is
 $\rho_{q2} = 0.625 \rho_{w},$ giving $\rho_{q2} = 0.75$ for $\rho_w = 1.2.$
The maximum wall distance for a RWTM, from \req{rhoq2}, for $\rho_{q2} = 1,$ is  $\rho_w = 1.5.$ 
Otherwise it is 
a no wall tearing mode 
\cite{nstx25}.

The importance of the $\rho_{q2}$ criterion is that the tearing mode 
behavior depends on the  wall boundary condition.
RWTMs grow to large amplitude, sufficient for a complete thermal quench.
If the wall is ideal, the modes only cause minor disruptions.
Feedback can emulate an ideal wall and prevent major disruptions.

It should be pointed out  that according to \rfig{sweeney}, 
most but not all disruptions satisfy the
two criteria, and not all shots that satisfy the criteria disrupt. These cases should
be investigated. 
  
\section{\large{\bf  DIII-D disruption simulations}} \label{sec:d3dsim}

The particular DIII-D shot 154576 was simulated \cite{d3d22}, initialized with an
equilibrium reconstruction. It was found that the equilibrium was linearly
stable, if the wall surrounding the plasma was ideally conducting 
and $q_0$ was slightly above unity.
It was unstable
if the wall was resistive. 
The linear growth rate  was shown to asymptotically scale as
$\gamma \propto \tau_{wall}^{-1},$  
the same  as an external kink resistive
wall mode (RWM) \cite{d3d22,mst23}, where $\tau_{wall}$  is defined after eq. \req{thin1}. 
 The location of the $q = 2$ radius   was  $\rho_{q2} \approx 0.8.$  
Nonlinear simulations reproduced the experimental timescale and gave a complete
thermal quench. 

\begin{figure}[h]
\hspace{.5cm}
\begin{center}
\includegraphics[height=4.5cm]{\figdir/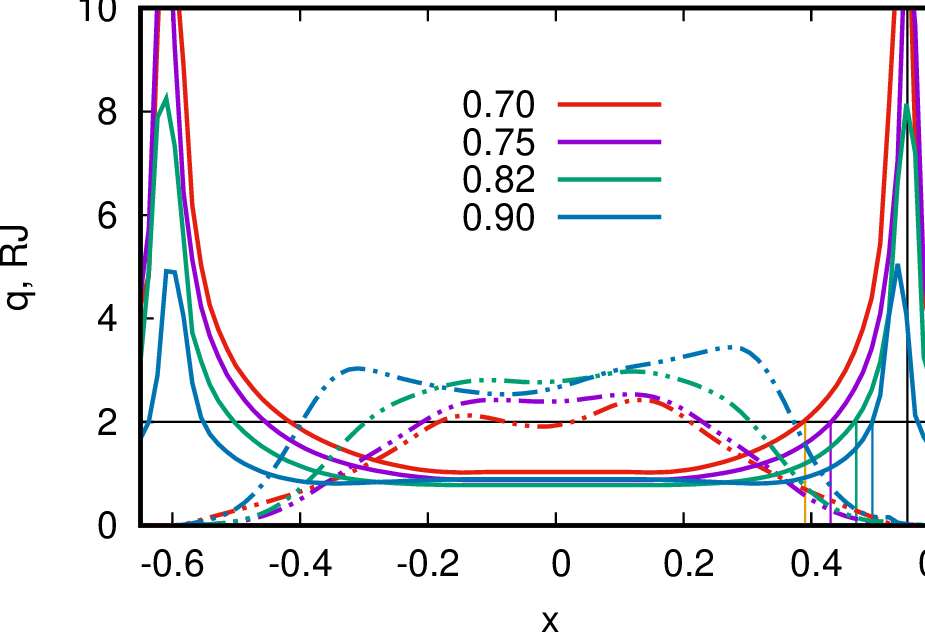}(a)
\includegraphics[height=4.5cm]{\figdir/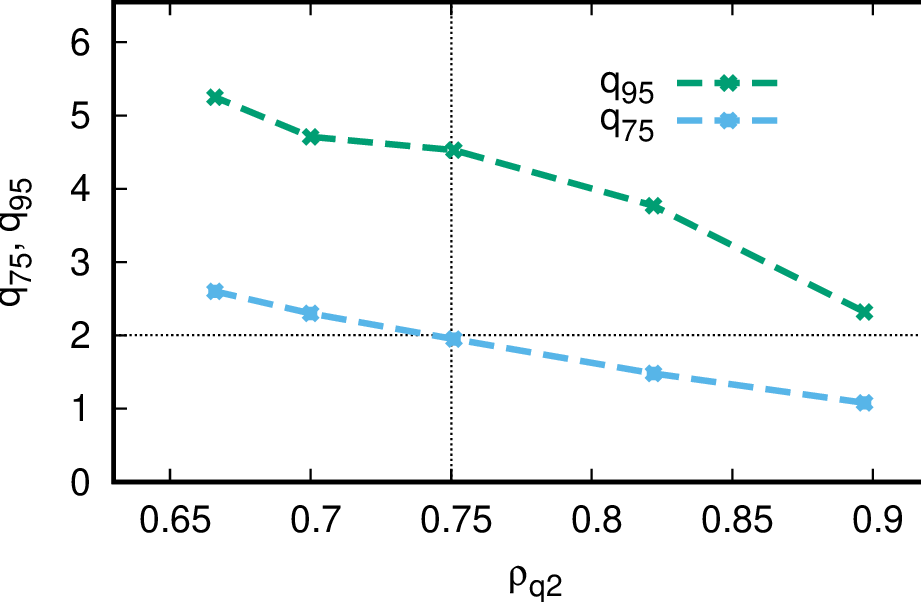}(b)
\end{center}
\caption{\it (a) $q$ and current density  $RJ$ as functions of $x = R - R_0$ in the
midplane. The curves are labelled with the values of $\rho_{q2}.$ 
(b) relation of $q_{75}$, $q_{95}$ to $\rho_{q2}.$}
\label{fig:d3dsim}
\end{figure}

To investigate the $\rho_{q2}$ criterion, 
a sequence of model equilibria was prepared with a method used for MST simulations \cite{nf-iaea24,model}.
Linear combinations of the current density in the  shot 154576 reconstruction  and
the square of the current density gave initial states in
which $q_0 = 1$ on axis and the total current was varied, giving a range of $\rho_{q2}$ values.
They have low $\beta_N.$ 
\rfig{d3dsim}(a) shows $q$ and current density  $RJ$ as functions of $x = R - R_0$ in the
midplane, where $R_0$ is the major radius.
The current density is small for $\rho \ge \rho_{q2}.$
The values of $\rho_{q2}$ are given as labels for the curves.  The  
values are $\rho_{q2} = x_{q2} / a,$
where $a  =0.55,$ and $x_{q2}$ is the value of $x$ for which $q = 2.$
\rfig{d3dsim}(b) shows $q_{75}$ and $q_{95}$ as  
a function of $\rho_{q2}.$ The quantity $q_{75}$ is a useful metric \cite{nstx25} for the
$\rho_{q2}$ criterion.

These states were used to initialize nonlinear M3D runs, with different boundary conditions.
The wall resistivity was either zero or such that $\tau_{wall} = 10^3 \tau_A,$ 
an order of magnitude larger than in  experiment. This was done to lower the
computation time. The other simulation parameters were the same as in  previous
simulations \cite{d3d22,nstx25}:  Lundquist number $S = 10^6$ with Spitzer like
temperature dependence, and parallel thermal conductivity $\kappa_\parallel = 10 R / v_A.$ 
The simulations were done using M3D \cite{m3d}.

\begin{figure}[h]
\vspace{.5cm}
\begin{center}
\includegraphics[height=4.2cm]{\figdir/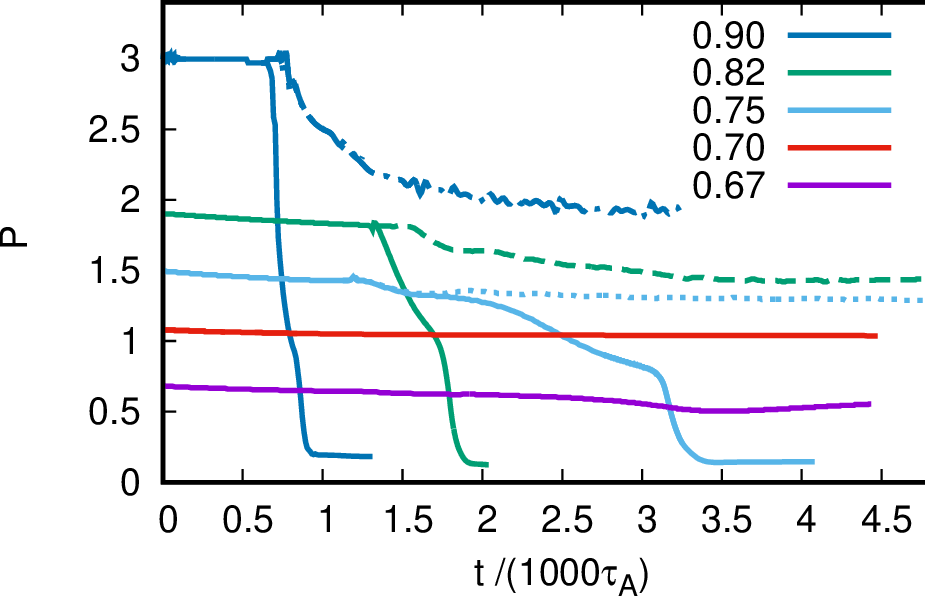}(a)
\includegraphics[height=4.2cm]{\figdir/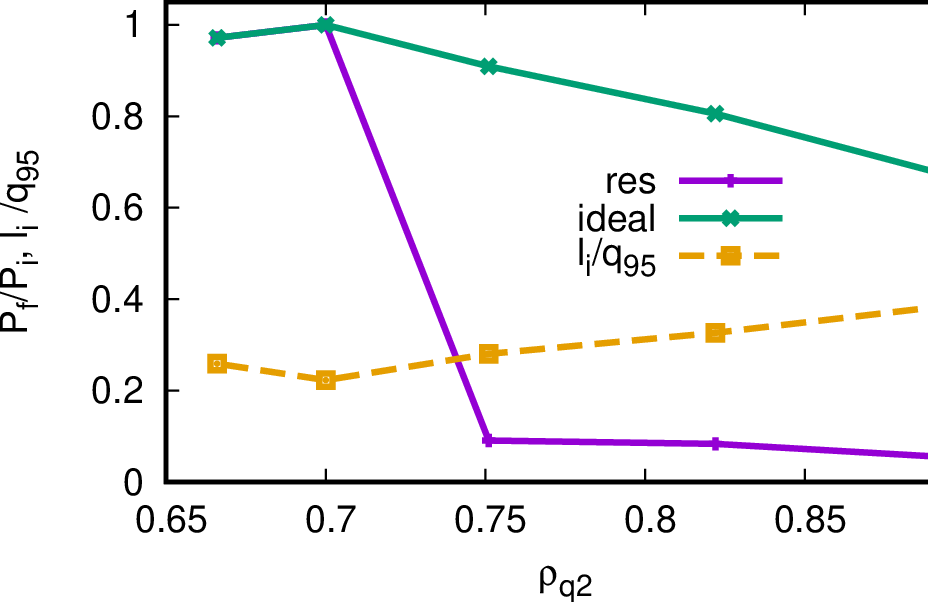}(b)
\end{center}
\caption{\it
(a) time history of total pressure $P$ for different boundary conditions.
Solid curves have a resistive wall, while dotted curves have an ideal wall.
(b) Nonlinear stability as a function of $\rho_{q2}$. 
$P_f/P_i$ is the ratio of final to initial total pressure $P$ in (a) 
as a function $\rho_{q2}.$ Also plotted is $l_i / q_{95}$ of the equilibria  in
\rfig{d3dsim}(a).}
\label{fig:simbase}
\end{figure}

Time history plots of nonlinear simulations  of 
these cases is shown in \rfig{simbase}(a),
which shows volume averaged  pressure
$P$ as a function of time for initial states with different $\rho_{q2}.$ 
The solid curves have a resistive wall,
while dotted curves have an ideal wall.
For  cases with $\rho_{q2} \ge 0.75,$ and a resistive wall,
a thermal quench occurred. With an ideal wall,
either  no thermal quench, or a minor quench,  occurred.

The minor disruptions in the time history plots of \rfig{simbase}(a)  are caused by
 $(1,1)$ modes. 
Interaction of $(1,1)$ and $(2,1)$ modes, presumably by toroidal coupling,
has been noted in experiments \cite{coupling5}. 
As seen in \rfig{d3dsim}(a), the $q = 1$ radius increases as $\rho_{q2}$ increases,
causing larger amplitude internal modes.
This effect could be prevented by keeping $q_0 > 1,$ which will not be examined here.

The results are summarized in \rfig{simbase}(b). 
The ratio of final to initial total pressure 
$P_f / P_i$ is
plotted as a function of $\rho_{q2}$, for both ideal and resistive wall boundary conditions.
In addition, the quantity $l_i /q_{95}$ of the equilibria  in \rfig{d3dsim}(a)  is plotted. 
\rfig{simbase}(b) contains the information  compared  with the data in \rfig{sweeney}
and \rfig{liq95}.


In the simulations, the dominant modes are $(m,n) = (2,1).$ 
The perturbed $\psi$ with resistive wall and no feedback,
for the case $\rho_{q2} = 0.82,$ is plotted in \rfig{lobe}(a) in the plane $\phi = 0,$
and in \rfig{d3dfbk}(a) for $\phi = \pi /2.$
The  pressure is shown  in \rfig{d3dfbk}(b), having undergone a thermal quench.

\section{\large{\bf  Feedback stabilization of RWTM}} \label{sec:feedback}
Feedback can emulate the effect of an ideally conducting wall.
Feedback experiments on DIII-D and RFX \cite{hanson,piovesan} showed stabilization of
RWMs with $q_{95}  \sim  2.$ Feedback was used to stabilize high $\beta$ RWM in
NSTX \cite{sabbagh2010},
DIII-D \cite{okabayashi2009}, 
and KSTAR \cite{kstar}. 
An attempt was made to control tearing modes in RFX  \cite{zanca2} 
with $q_a > 2$, but encountered density limit disruptions.
\begin{figure}[h]
\begin{center}
\includegraphics[height=4.25cm]{\figdir/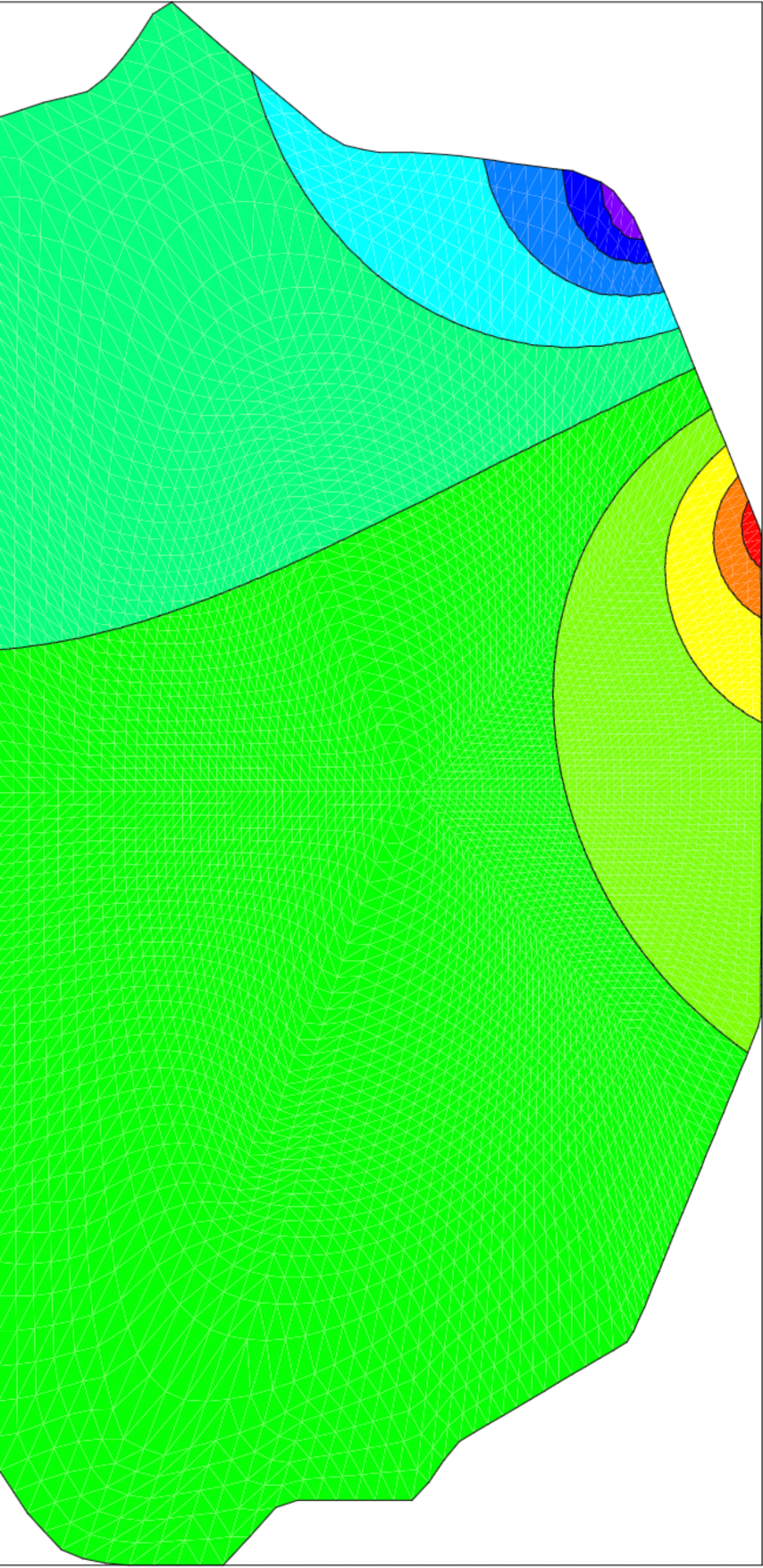}(a)
\includegraphics[height=4.25cm]{\figdir/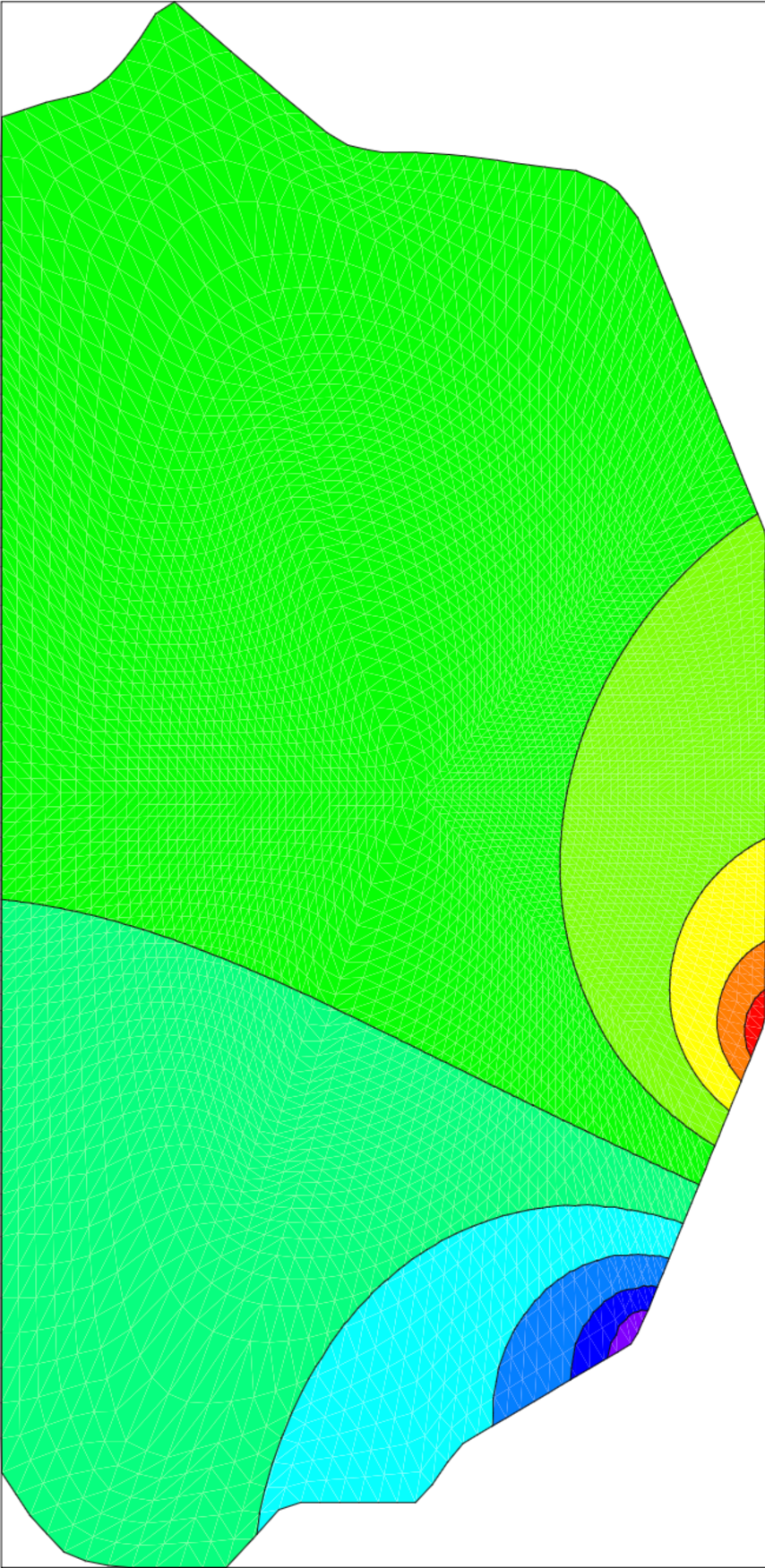}(b)
\includegraphics[height=4.25cm]{\figdir/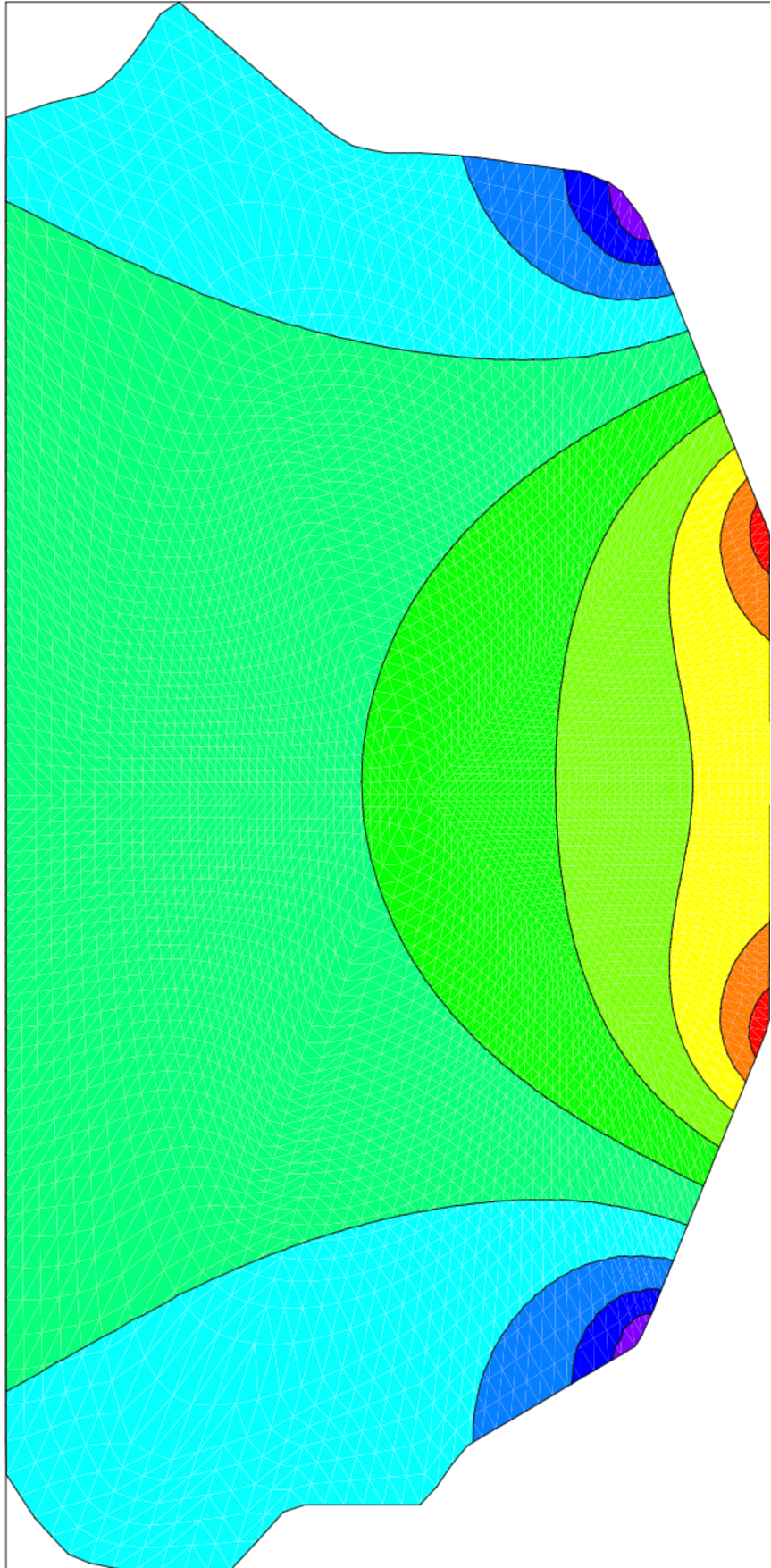}(c)
\includegraphics[height=4.25cm]{\figdir/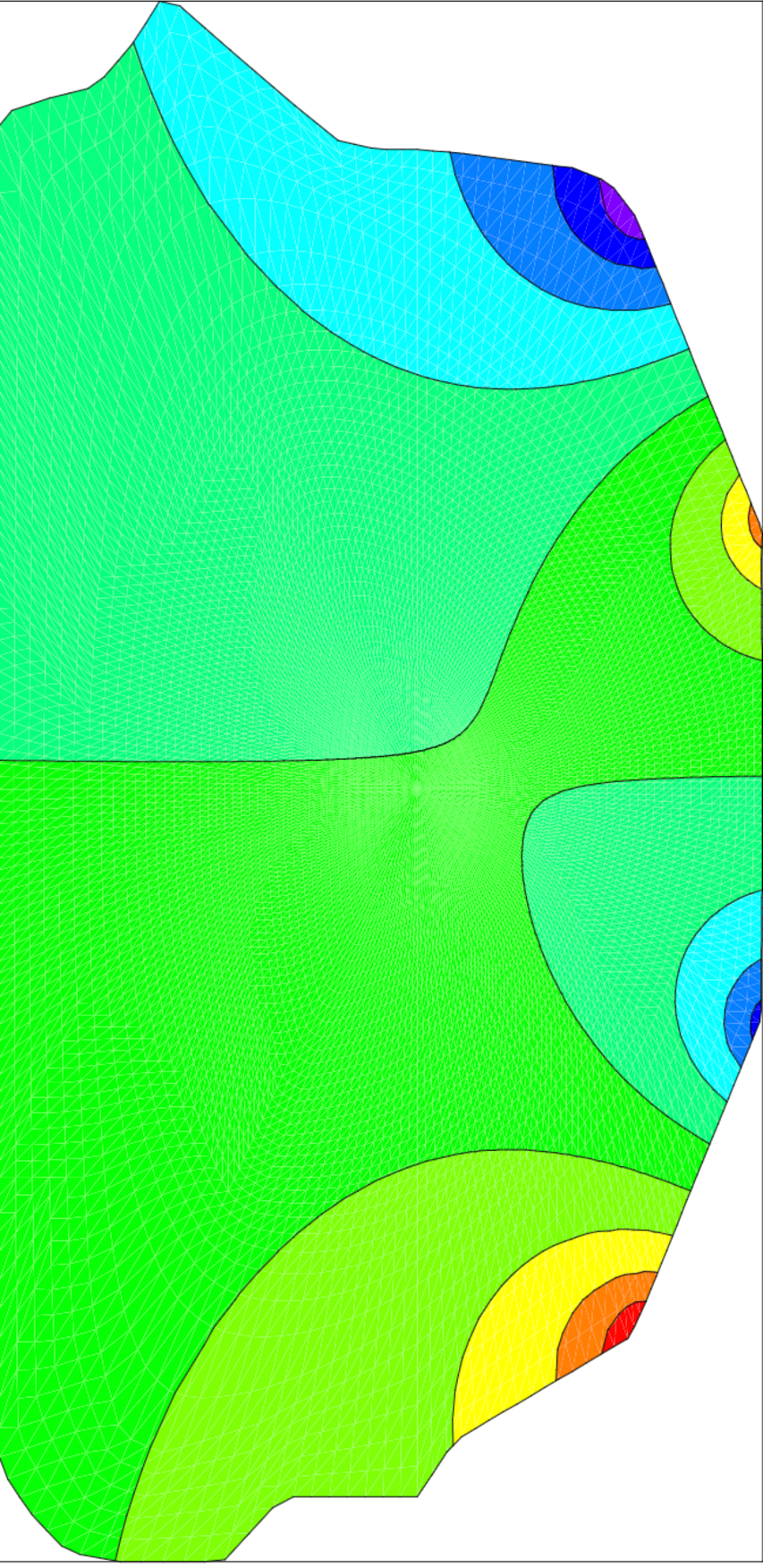}(d)
\end{center}
\caption {\it (a) $\Psi_U$ produced by  upper   I-coil; 
           (b) $\Psi_L$ produced by  lower I-coil; 
           (c) $\Psi_U + \Psi_L$ ; 
           (d) $\Psi_U - \Psi_L$ . 
The coil configuration (c) in $\psi_{f1}$ of eq. \req{sens1}  is effective for stabilizing (2,1) RWTMs, while (d) by itself is ineffective. } 
\label{fig:coils}
\end{figure}

In DIII-D, feedback was provided using the I-coils, sets of window pane type coils just outside
the wall.  Here 
feedback is modeled using 
the thin wall boundary condition \cite{jet21,d3d22,finn95,nf-iaea24},
\be \frac{\pd \psi_w}{\pd t} = \frac{r_w}{\tau_{wall}} ( \psi'_{vac} -  \psi_p' ) %
 - \gamma_f \psi_{fbk} 
\label{eq:thin1} \ee
where
$\tau_{wall} = \mu_0 r_w \delta_w / \eta_w,$
$\eta_w,\delta_w$ are wall resistivity and thickness, $r_w = R_w - R_0,$
the difference of wall major radius and magnetic axis, 
$\psi_w$ is the magnetic flux at the wall,
$\psi'_{vac}$ is the vacuum magnetic flux normal derivative at the wall excluding the
contribution of the feedback current, and $\psi_p'$ is the magnetic flux normal derivative
from the plasma at the wall.
In  previous studies \cite{nstx25,nf-iaea24}, $\psi_{fbk}$ was simplified to $ \psi_w.$  

\begin{figure}[h]
\begin{center}
\includegraphics[height=3.25cm,width=2.5cm]{\figdir/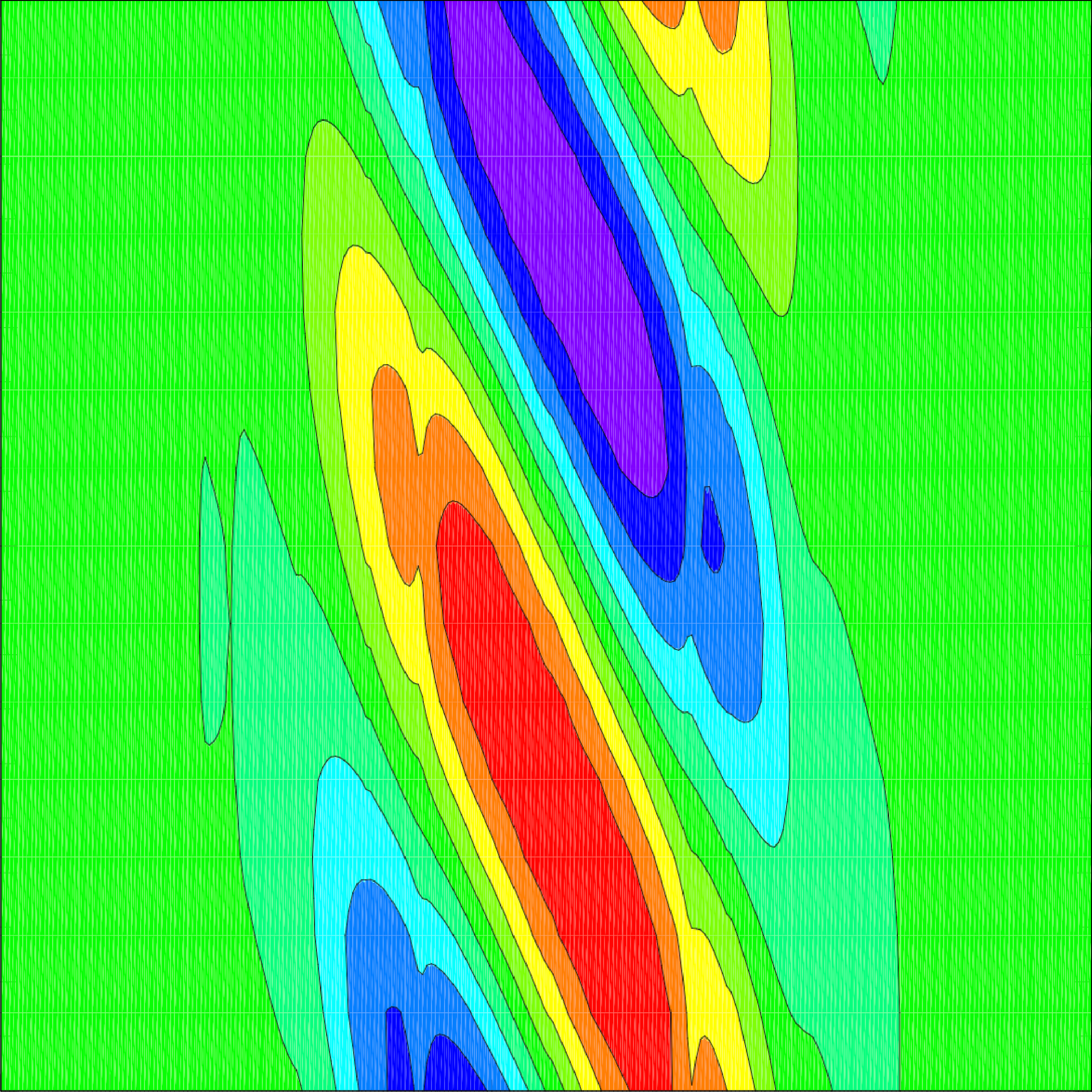}(a)
\includegraphics[height=3.25cm,width=2.5cm]{\figdir/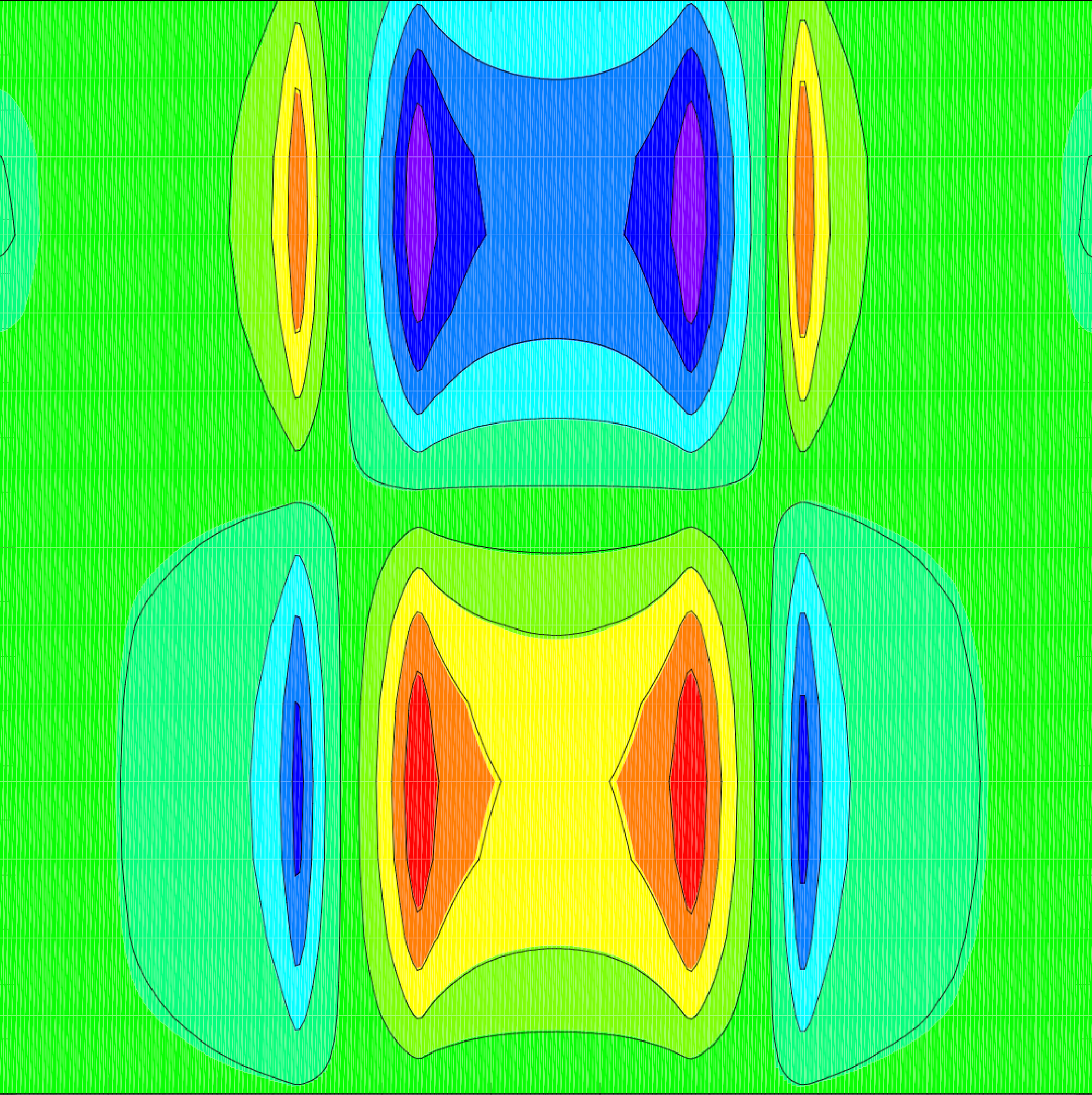}(b)
 \includegraphics[height=3.25cm,width=2.5cm]{\figdir/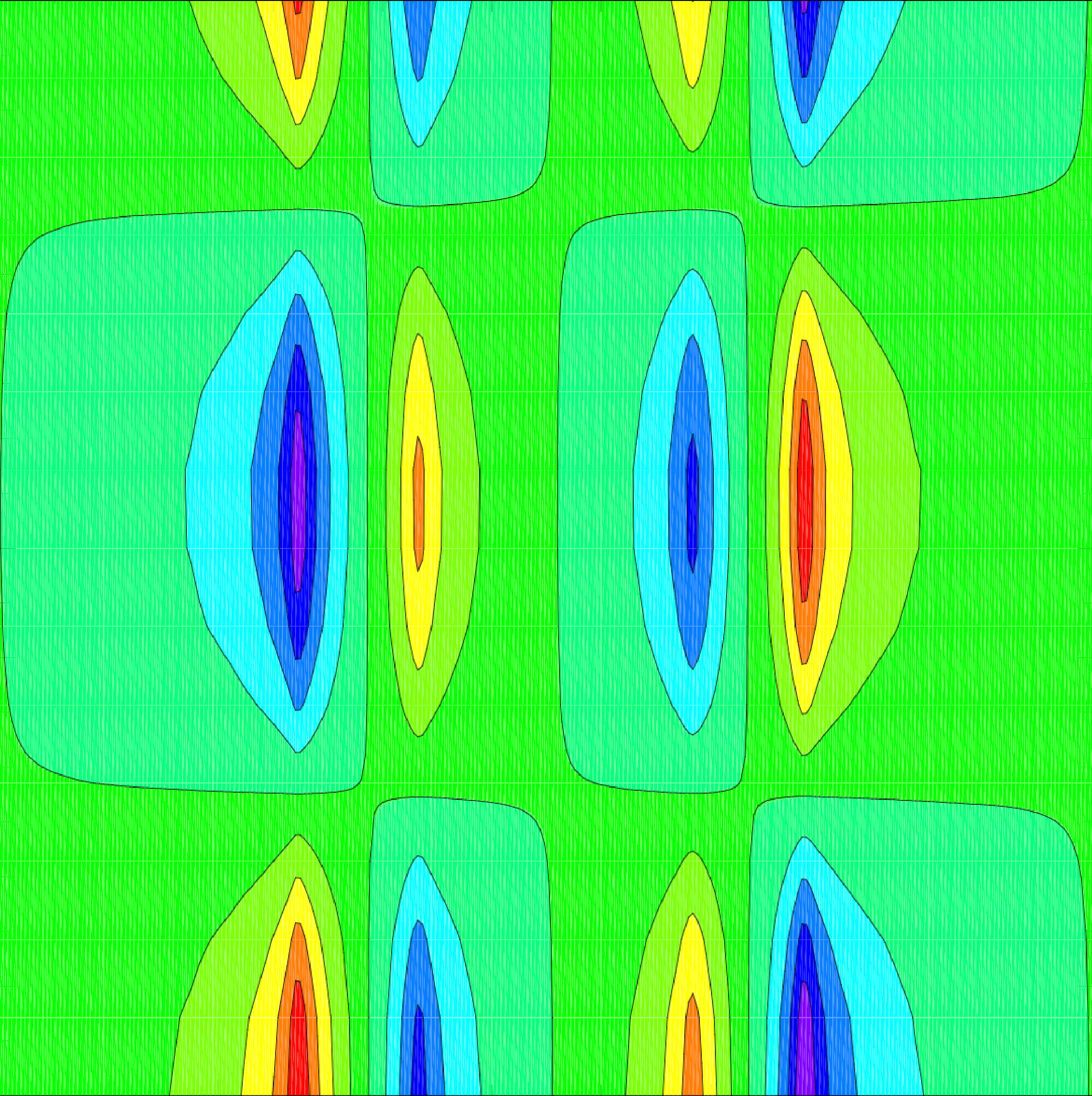}(c)
\includegraphics[height=3.25cm,width=2.5cm]{\figdir/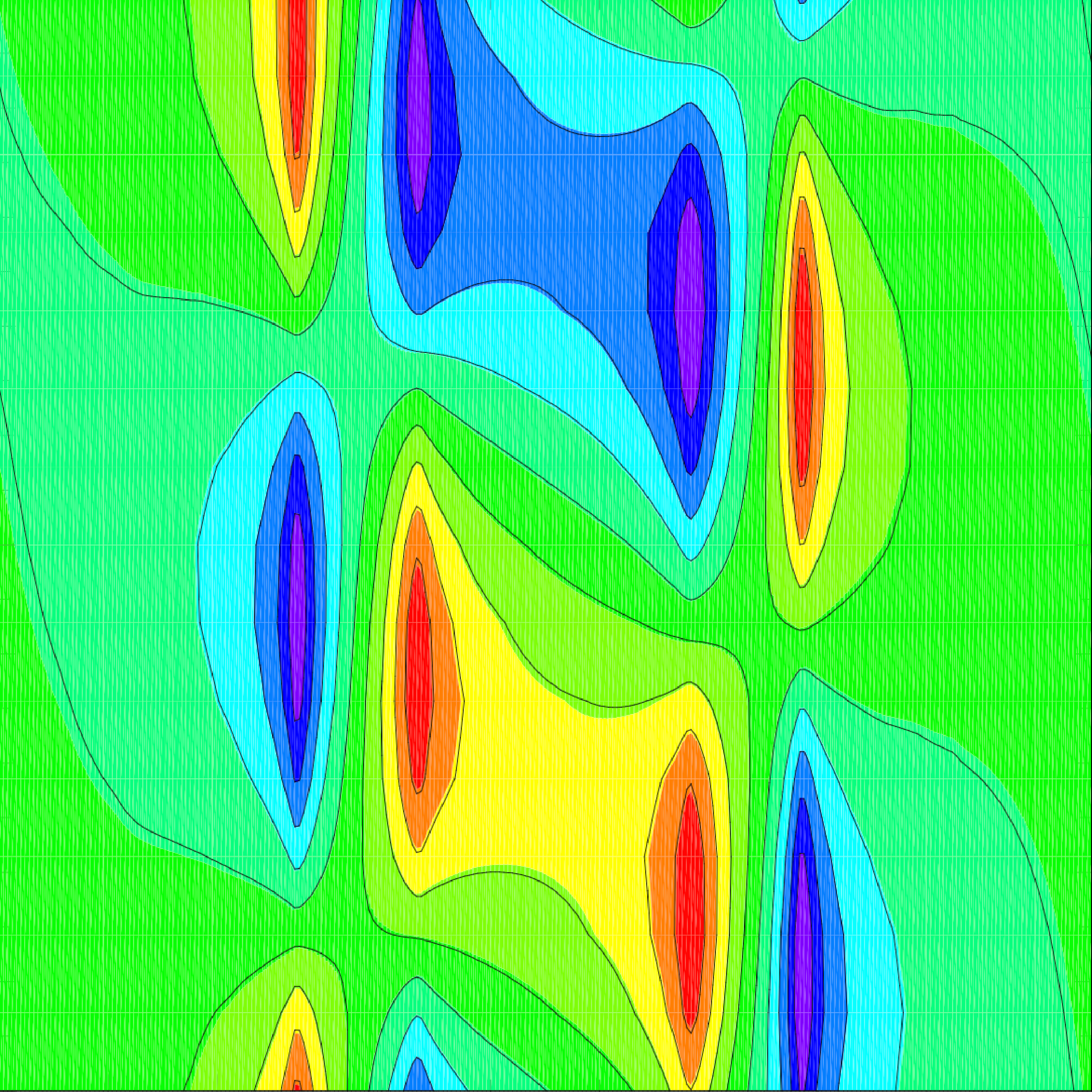}(d)
 \includegraphics[height=3.25cm,width=2.5cm]{\figdir/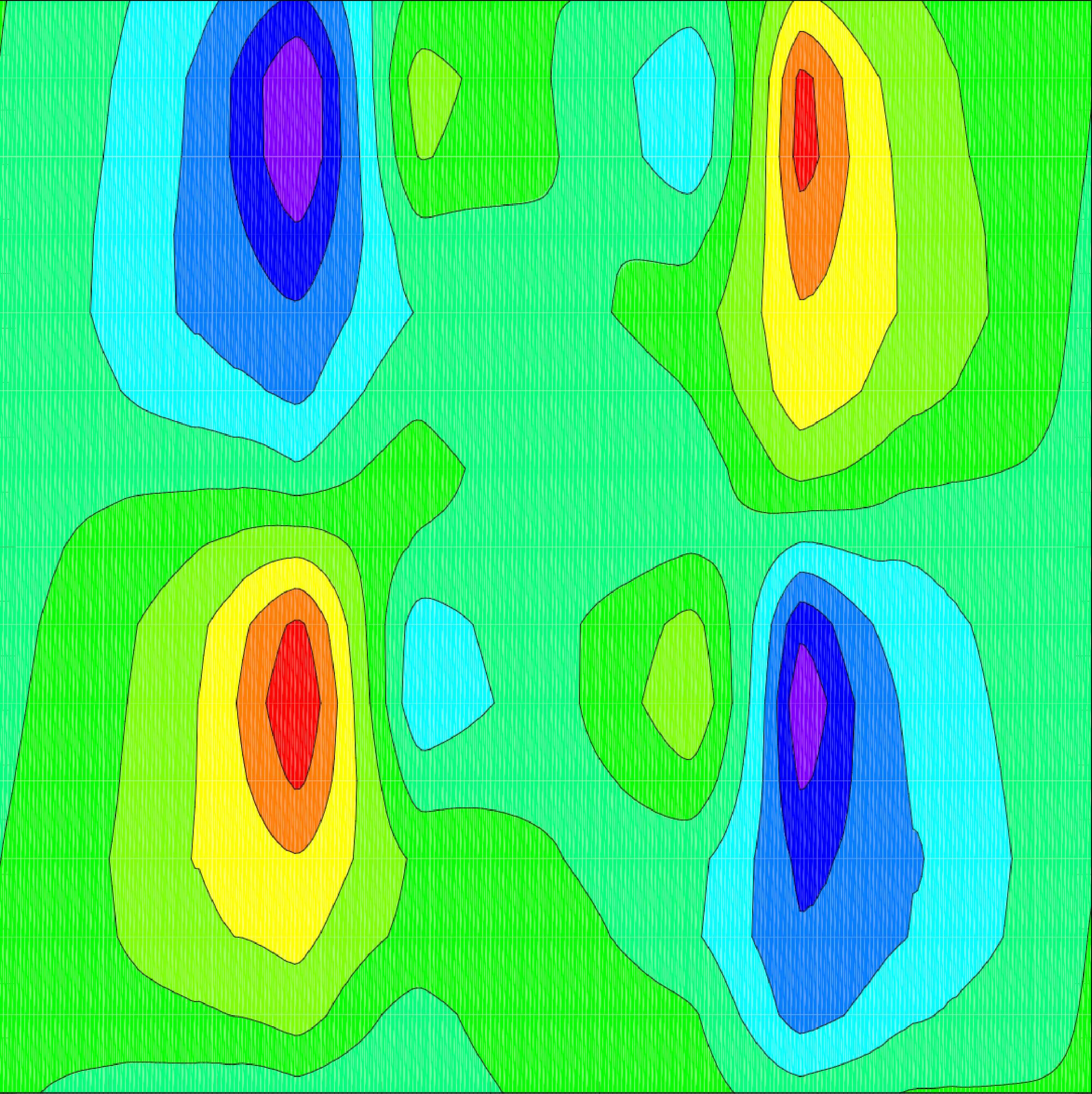}(e)
\end{center}
\caption {\it  (a) perturbed $\psi_w$, where  the horizontal 
coordinate is the 
poloidal distance along the wall, 
centered at the outer midplane. The vertical 
coordinate is the toroidal angle $\phi.$ 
 It is the  case shown in \rfig{lobe}(a),\rfig{d3dfbk}(a), with no feedback.
 (b) Feedback $\psi_{f1}$ in eq. \req{sens1};
 (c)  $\psi_{f2}$;
 (d)  $\psi_{f1}+\psi_{f2}$;
 (e) perturbed $\psi_w$ for resistive wall with only $\psi_{f2}$  feedback.
Poloidal perturbations $\psi(R,Z,\pi/2)$ are shown in \rfig{d3dfbk}(c). 
}
\label{fig:d3dwall}
\end{figure}

The flux on the wall from each pair of coils is of the form
$\psi_{coil}  \propto I_{coil}(\phi) \Psi_{coil}(R,Z),$ where the coil current $I_{coil}$ 
is a function of
toroidal angle. A continuous rather than discrete  model of $I_{coil}$ angular dependence is  
used \cite{fitzpatrick98,mars}.  
The coil current is proportional to $\psi_w$ sensors on the wall along the midplane.
Because $\psi_w$ has predominantly $n = 1$ toroidal
mode number, the discrete effects involving higher mode numbers are assumed small.

The flux  $\Psi_{coil}$ produced by the coils is
approximately  the flux from  pairs of  parallel poloidal coils
outside the resistive wall, which in DIII-D are the I-coils. 
The flux from a single  coil  is approximately  $\Psi  \propto \ln r,$ 
where $r = 0$ at the poloidal location of the coil.  
The flux $\Psi_U,\Psi_L$ from the upper and lower  pairs of coils, 
having  equal and opposite currents within each pair,  is approximately 
\be \Psi_{U}  =  
 \Psi_0 \ln\left[\frac{(R - R_{UL})^2 + (Z - Z_{UL})^2}
          {(R - R_{UU})^2 + (Z - Z_{UU})^2} \right] 
\label{eq:ws1} \ee
\be 
\Psi_{L}  = 
 \Psi_0 \ln\left[\frac{(R - R_{LL})^2 + (Z - Z_{LL})^2}
          {(R - R_{LU})^2 + (Z - Z_{LU})^2} \right] 
\label{eq:ws} \ee
where the upper coils are located 
at $R_{UU},Z_{UU}$ and $R_{UL}, Z_{UL},$ with $Z_{UU} > Z_{UL}$, 
and the lower coils are at
$R_{LL} = R_{UU}$, $R_{LU} = R_{UL},$ $Z_{LL} = -Z_{UU}$, $Z_{LU} = - Z_{UL}.$  
The constant $\Psi_0$ is chosen to normalize the maximum values of $ \Psi_{U},\Psi_{L}$
on the wall to unity.
The feedback flux $\psi_{fbk}$ is  a sum of $\Psi$ values weighted by sensor
values of $\psi_w$ on the wall outer midplane.

Suppose a $(2,1)$  RWTM varies on the wall approximately as $\psi_w(\theta,\phi) 
\sim  \cos(2\theta - \phi)$ $ = \cos(2\theta)\cos\phi + \sin(2\theta)\sin\phi.$
To approximate $\psi_{fbk} \sim \psi_w,$ using values of $\psi_w(0,\phi)$
at the midplane $\theta = 0,$
note that $\psi_w(0,\phi) = \cos\phi,$ and
$\psi_w(0,\phi+ \pi/2) - \psi_w(0,\phi - \pi /2) = -2 \sin\phi.$
Qualitatively near $\theta = 0,$ $\cos(2\theta) \sim (\Psi_U + \Psi_L),$
and $\sin(2\theta) \sim (\Psi_U - \Psi_L).$
Combining gives a possible choice for $\psi_{fbk}$,
\be \begin{array}{lcl} 
 \psi_{fbk} & = & \psi_{f1} + \psi_{f2}  \\
\psi_{f1} & = & \psi_w(0,\phi)\left(\Psi_{U}+\Psi_{L}\right)  \\
 \psi_{f2} & = &  \frac{1}{2}
\left[\psi_w(0,\phi+\pi/2) 
- \psi_w(0,\phi-\pi/2)\right]\left(\Psi_{L}-\Psi_{U}\right)  \end{array} \label{eq:sens1}  \ee
The alignment of the feedback $\psi_{fbk}$  with the mode $\psi_w$  
 in principle can be optimized \cite{mars2} for every equilibrium and $\rho_{q2}.$
This will not be considered here.

The $\psi_{fbk}$ perturbations are displayed in \rfig{d3dwall}, 
where (a) is the perturbed
$\psi_w$ in \rfig{lobe}(a), \rfig{d3dfbk}(a), with time history in 
\rfig{simbase}(a) with $\rho_{q2} = 0.82,$ at time $t = 1342 \tau_A,$ with
resistive wall and no feedback. In the plot, the horizontal axis is poloidal length,
centered at the outer midplane, and the vertical axis is toroidal angle $\phi.$ 
\rfig{d3dwall}(b) is $\psi_{f1}$, (c) is  $\psi_{f2}$,
and (d) is $\psi_{f1}+\psi_{f2}.$
Although $\psi_{f2}$ is initially stabilizing, it allows the development of
an unstable feedback induced mode, 
shown in \rfig{d3dwall}(e),  in \rfig{d3dfbk}(c)
and in the time histories of \rfig{d3dsim2}(a),(b).
This mode does not occur if $\psi_{f1}$ or $\psi_{f1}+\psi_{f2}$ is used  for feedback,
as shown in \rfig{d3dsim2}(a),(b).

To assess the effectiveness of $\psi_{fbk}$, 
it is useful to simplify   \req{thin1} to 
\be \frac{1}{2}\frac{\pd <\psi_w^2>}{\pd t} \approx \gamma <\psi_w^2> - \gamma_f <\psi_w\psi_{fbk}>
= \left(\gamma - {\gamma_f}{C_{fbk}} \right) <\psi_w^2> \label{eq:gammacor} \ee
where $\gamma$ is the RWTM growth rate without feedback, and $\psi_w$ in  the first term on the
right side of \req{gammacor} is substituted for the first terms on the right of \req{thin1}.
The correlation of $\psi_w$ and $\psi_{fbk} $ is
\be C_{fbk} = \frac{<\psi_w\psi_{fbk}>}{< \psi_{w}^2>}
 \label{eq:cor} \ee
where $< .. > = \oint\oint .. dl d\phi,$ and $dl$ is a length element tangent to the wall.
Correlations  are listed in \rtab{qart}, between $\psi_w$ and $\psi_f$ in
\rfig{d3dwall}.
\begin{table}[h]
\begin{center}
\begin{tabular}{|r|r|r|}
\hline
$\psi_w$  & $\psi_{f}$ & $C_{fbk}$   \\
\hline
a & b & $0.331$   \\
a & c & $0.223$   \\
e & b & $0.044$   \\
e & c & $-0.67$   \\
\hline
\end{tabular}
\end{center}
\caption{\it Correlation of $\psi_w,$ and $\psi_{fbk}$  in \req{cor}.
The first column is $\psi_w$ in \rfig{d3dwall}(a),(d), and the second column
is $\psi_{fbk}$ in \rfig{d3dwall}(b),(c). The correlations are in the third column.}
\label{tab:qart}
\end{table}
It can be seen from \rtab{qart}  that the choice 
$\psi_{fbk} = \psi_{f1}$ is always stabilizing. 
The case \rfig{d3dwall}(d),$\psi_{f2}$ has a negative
correlation, so that the feedback destabilizes the mode.

If $C_{fbk} > 0,$ then the mode is suppressed if
\be \gamma_f > \frac{\gamma}{C_{fbk}} \label{eq:gammaf} \ee
If $\gamma \approx 4 / \tau_{wall}$ \cite{mst23}, and $C_{fbk} = 1/3$ from \rtab{qart},
 then $\gamma_f \stackrel{>}{\sim}  12 / \tau_{wall}$ should be sufficient  to
prevent a  thermal quench.
The value of $\gamma_f = 0.05$ in all the simulations, not taking into account 
the $C_{fbk}$ variation in \rtab{qart}.  
\section{\Rdd{  DIII-D  feedback simulations }} \label{sec:d3d}
\begin{figure}[h]
\vspace{.5cm}
\begin{center}
\includegraphics[height=4.2cm]{\figdir/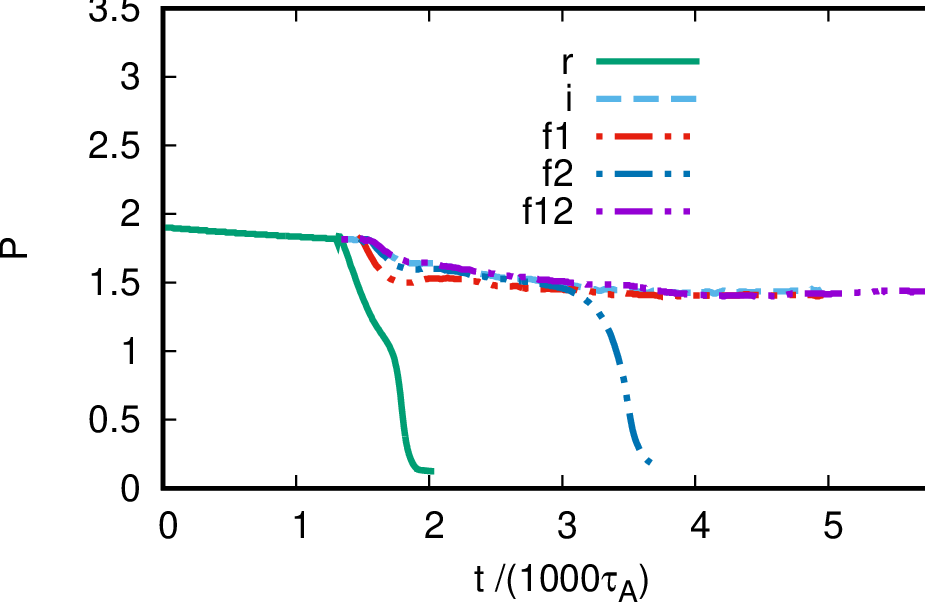}(a)
\hspace{.2cm}
\includegraphics[height=4.2cm]{\figdir/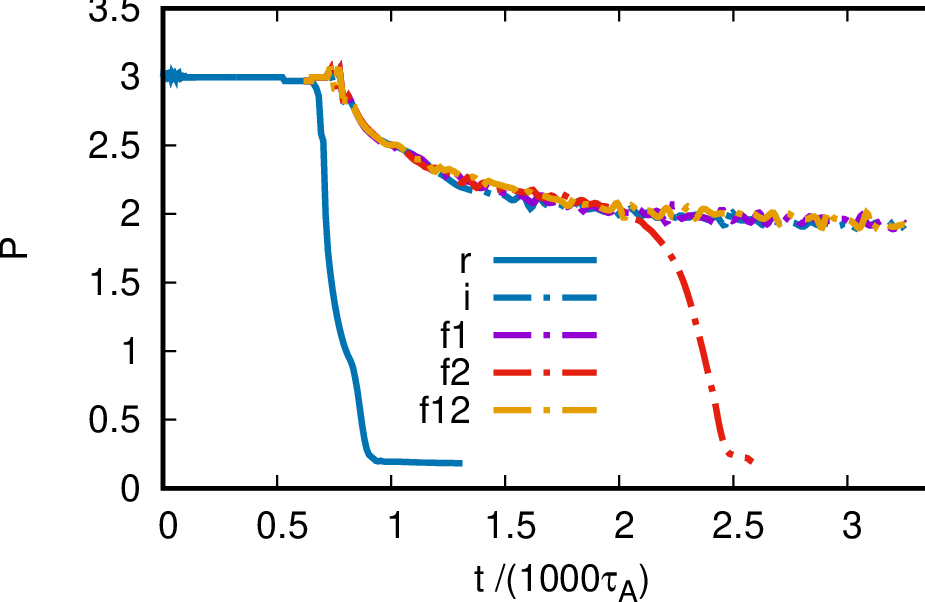}(b)
\end{center}
\caption{\it 
 time history of total pressure $P$ for different boundary conditions and feedback.
``r" is resistive wall with no feedback; ``i" is ideal wall;
``f1" is $\psi_{fbk}= \psi_{f1}$; ``f2" is $\psi_{f2}$; and ``f12" is
$\psi_{fbk} = \psi_{f1}+ \psi_{f2}$.
(a) $\rho_{q2} = 0.82$. 
(b)  $\rho_{q2} = 0.90$.
}
\label{fig:d3dsim2}
\end{figure}

\begin{figure}[h]
\begin{center}
\includegraphics[height=4.2cm]{\figdir/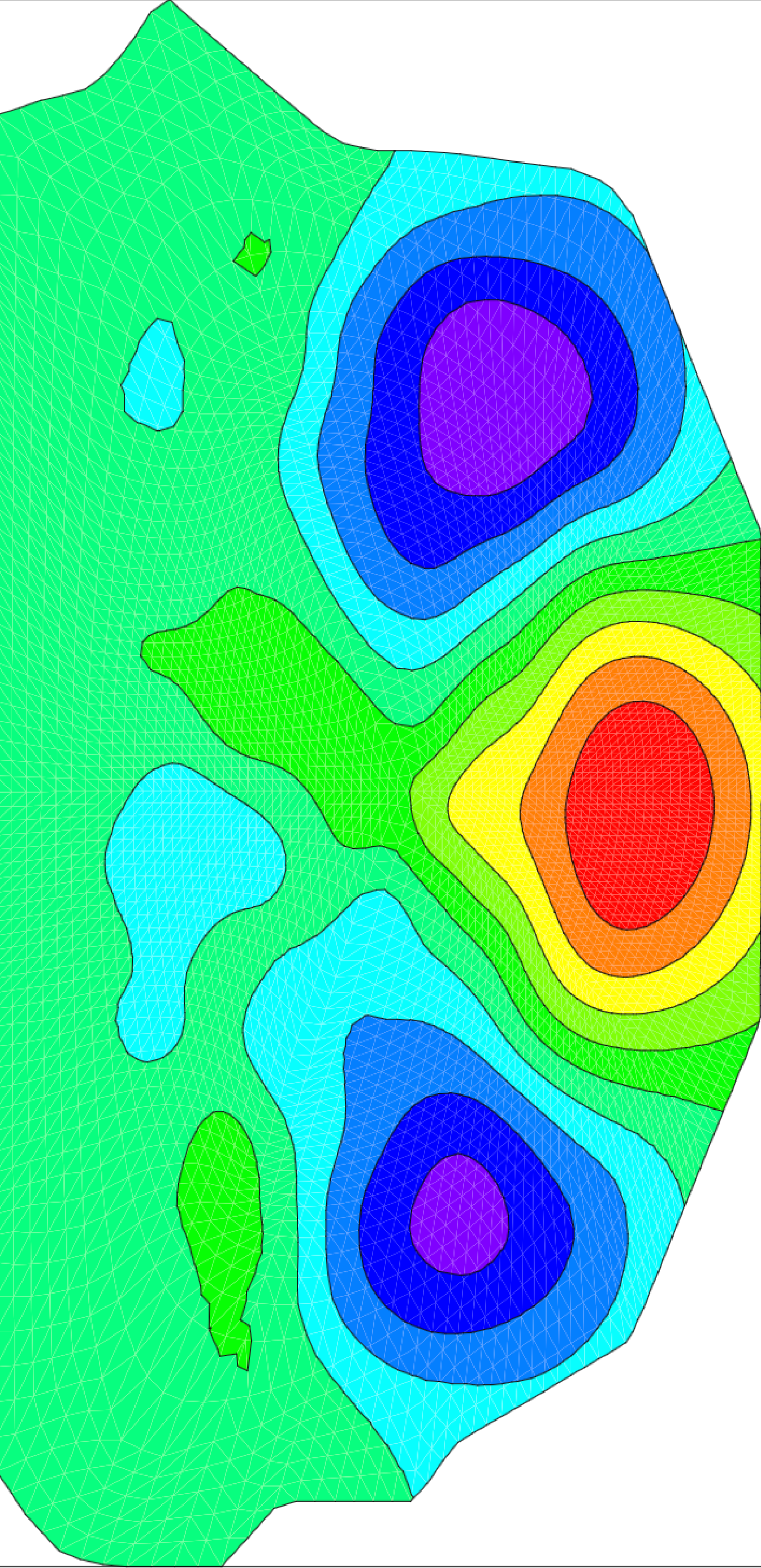}(a)
\includegraphics[height=4.2cm]{\figdir/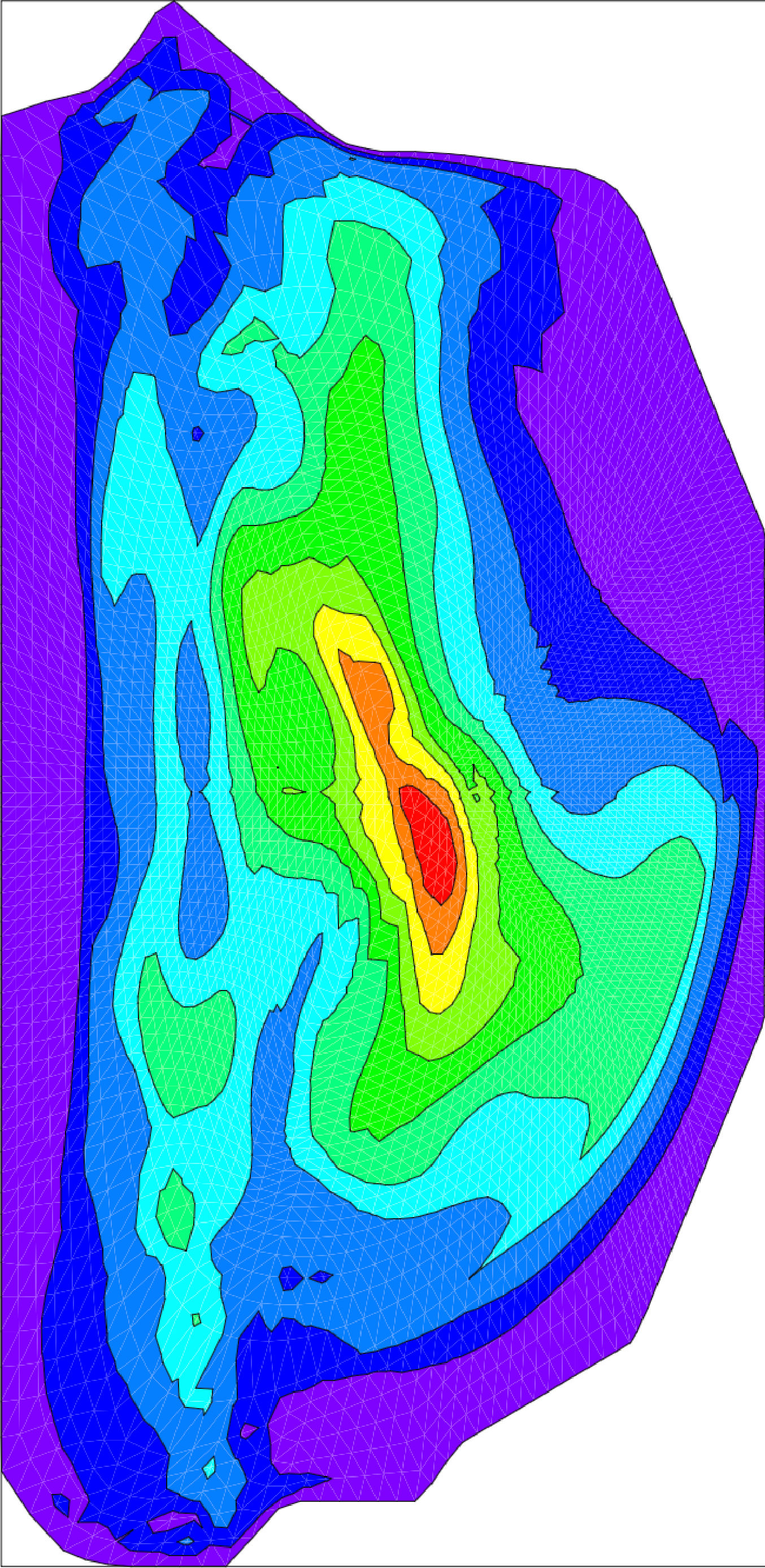}(b)
\includegraphics[height=4.25cm]{\figdir/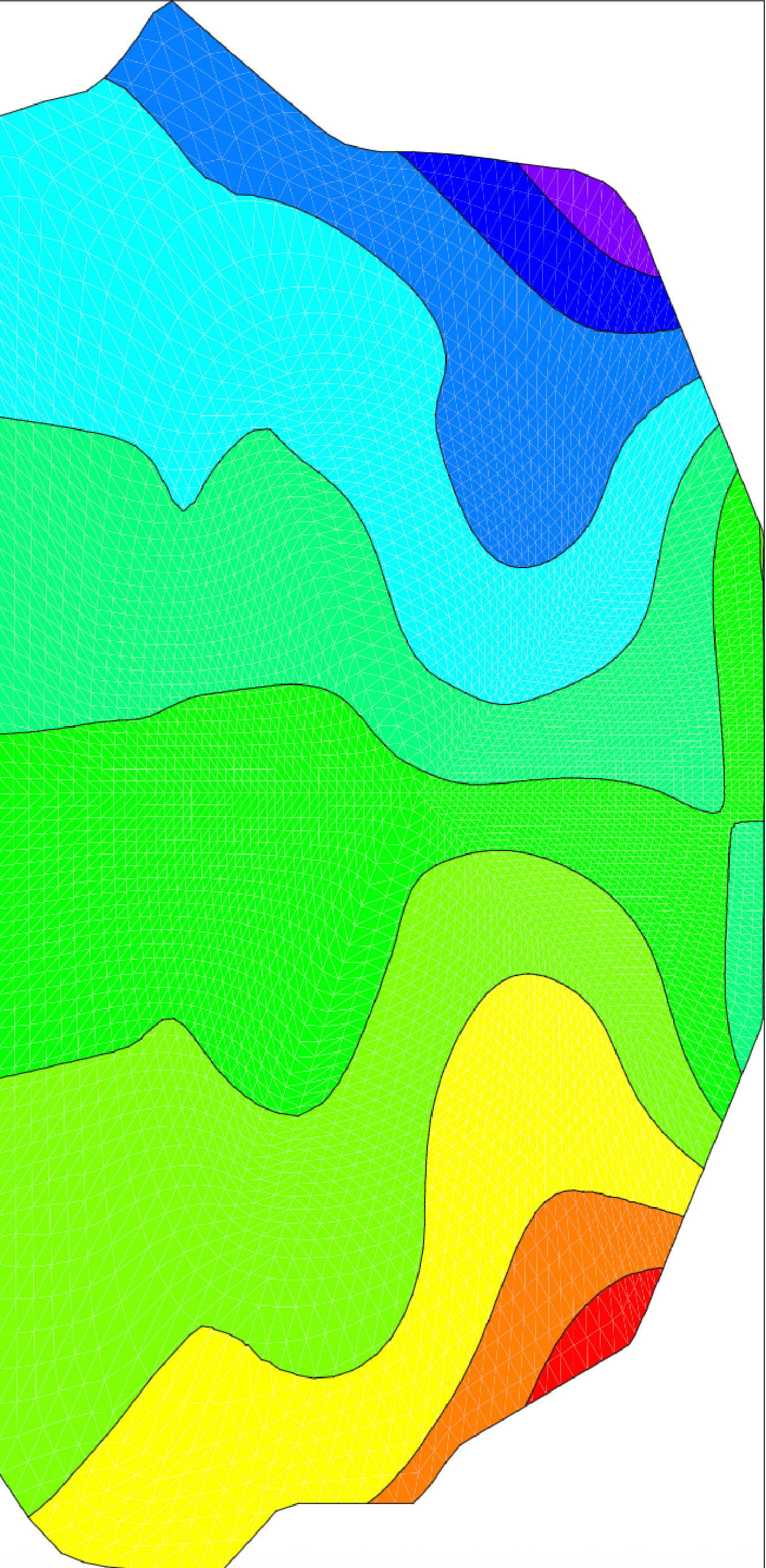}(c)
\includegraphics[height=4.25cm]{\figdir/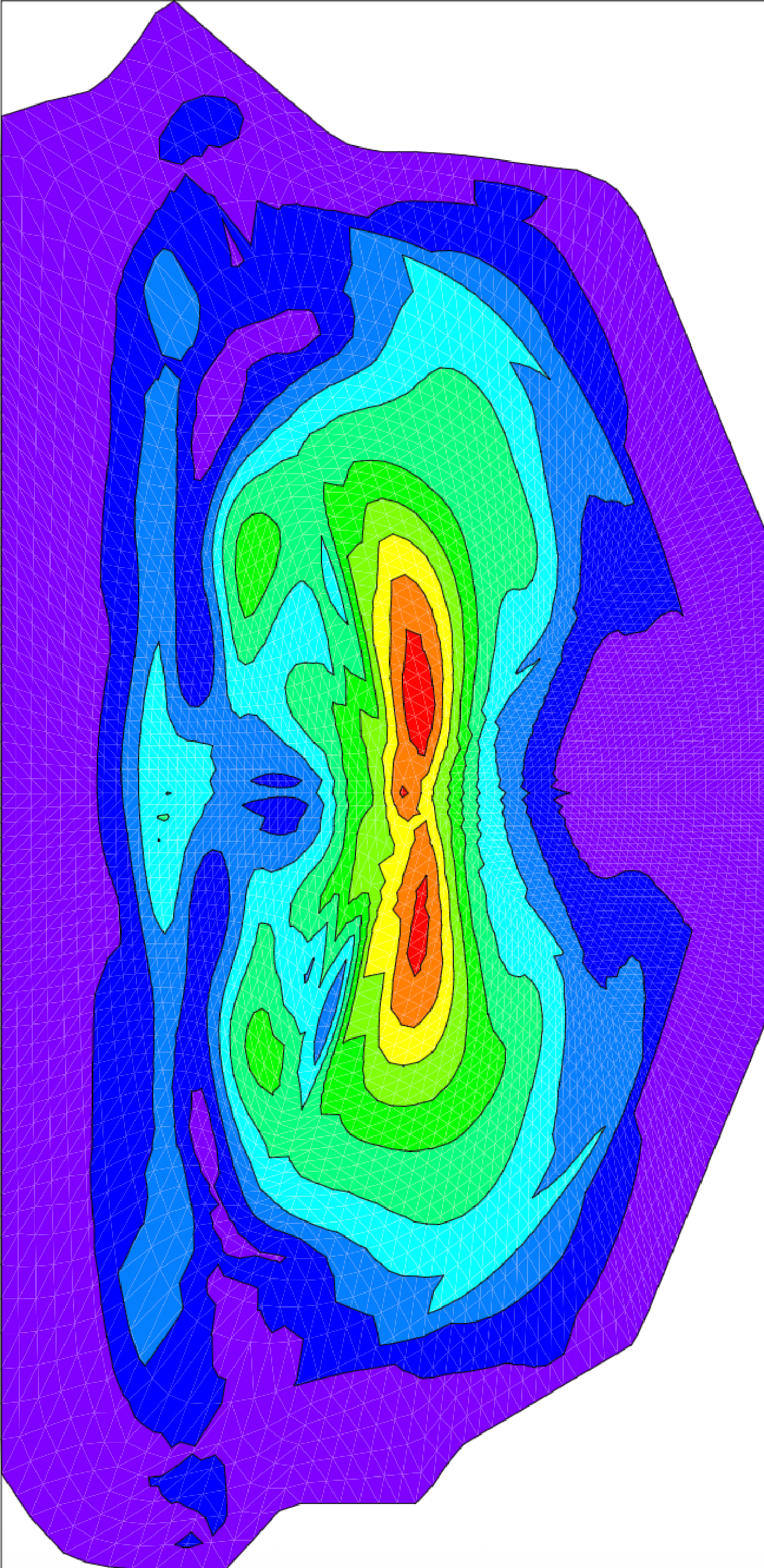}(d)
\includegraphics[height=4.25cm]{\figdir/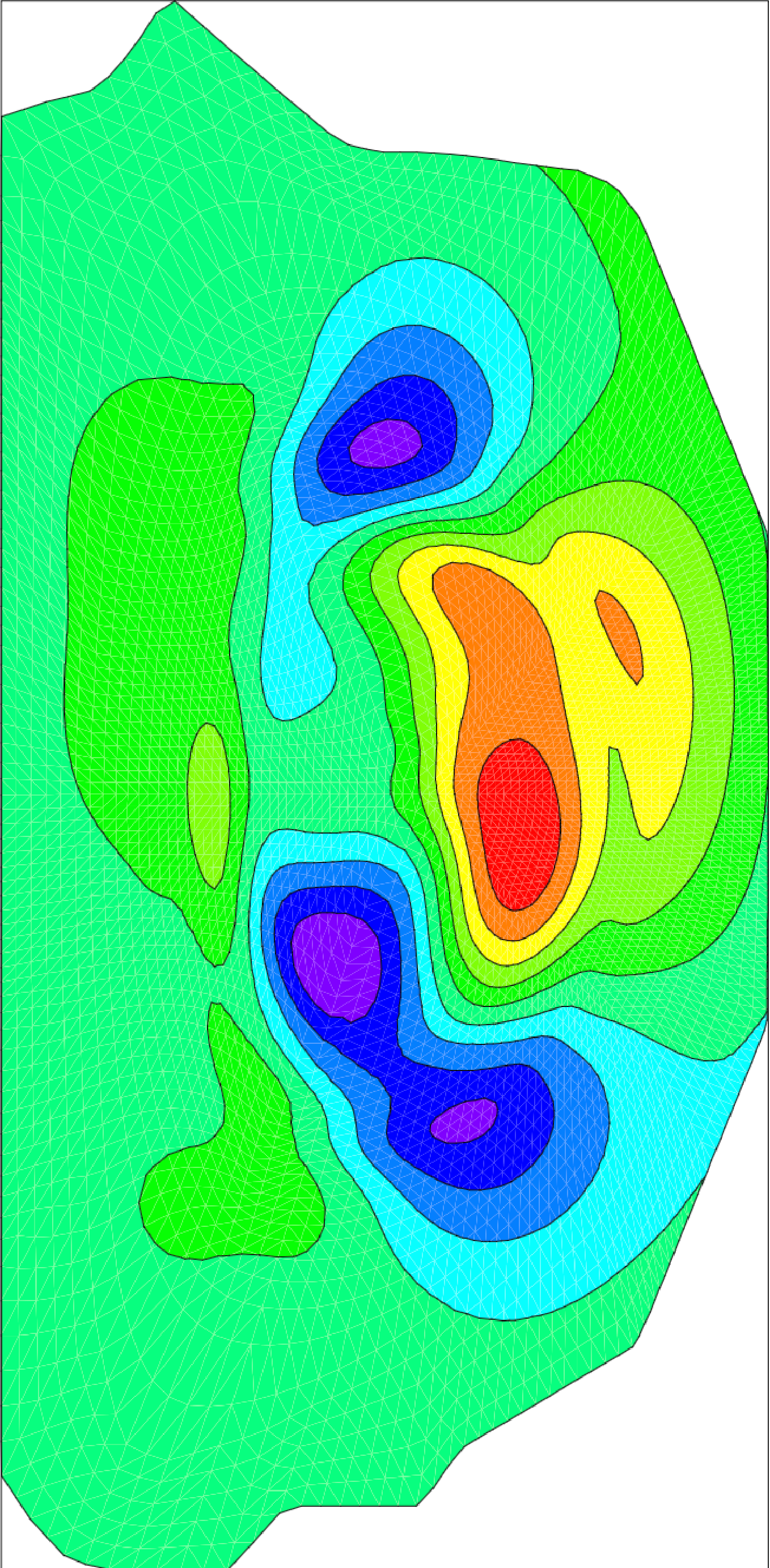}(e)
\includegraphics[height=4.25cm]{\figdir/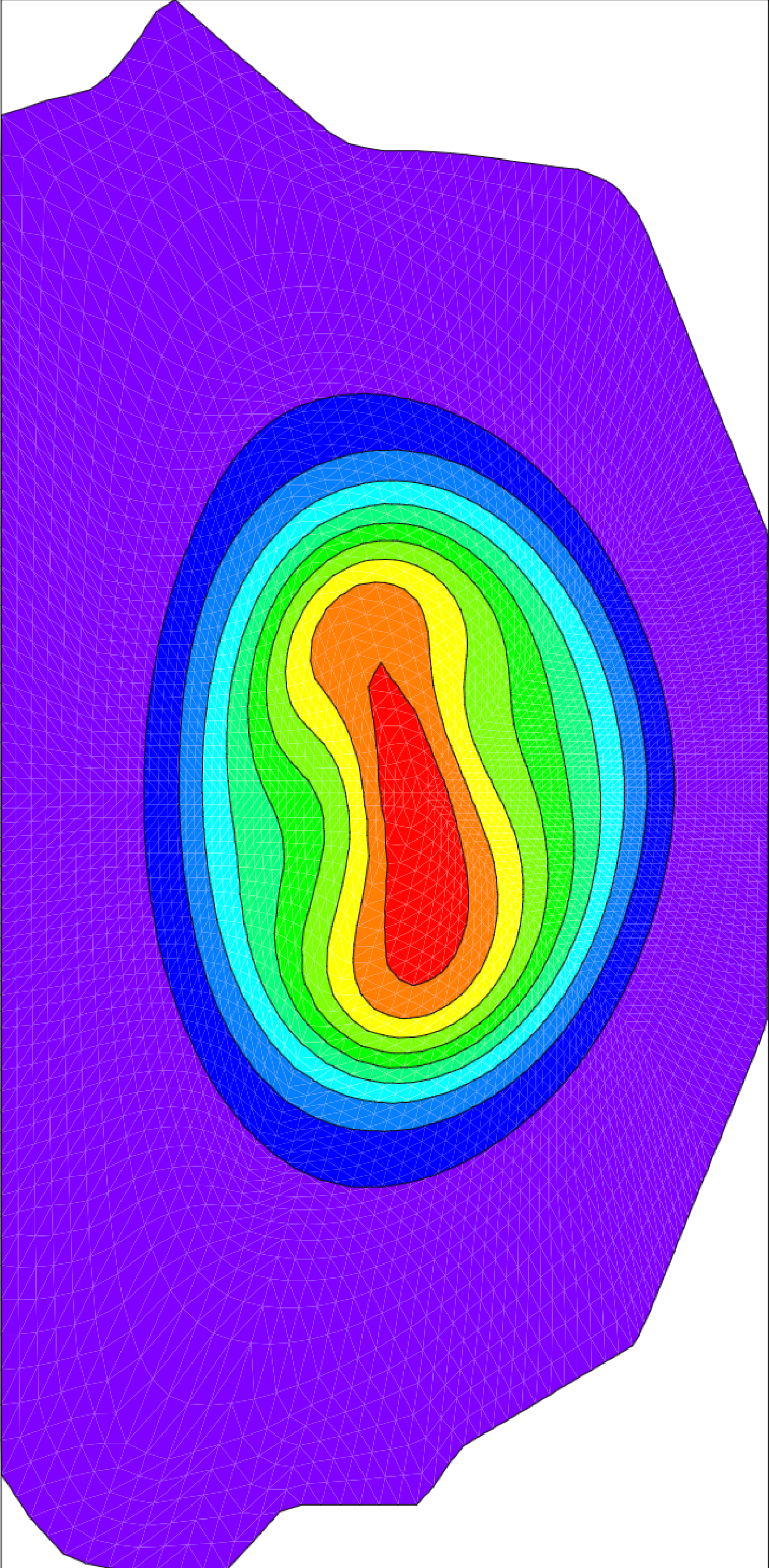}(f)
\end{center}
\caption {\it  Cases with resistive wall: (a) perturbed $\psi$, no feedback; (b) pressure $p$, 
both at time 
$t = 1820 \tau_A,$ labeled ``$0.82$" in \rfig{simbase}(a), and  ``r" in \rfig{d3dsim2}(a);
 (c) perturbed $\psi$ with feedback using $\psi_{f2}$, labeled ``f2" in  \rfig{d3dsim2}(a) at
time $t = 3657 \tau_A$; (d)  pressure $p$ for same case; 
(e) $\psi$ with feedback using $\psi_{f1}$ in
case labelled ``f1" in  \rfig{d3dsim2}(a) at  time $t = 4059 \tau_A;$
 (f) $p$ for same case. }
\label{fig:d3dfbk}
\end{figure}

In \rsec{d3dsim} simulations were done with ideal and resistive wall boundary
conditions.
Here feedback $\psi_{fbk}$ 
is added at the resistive wall.
\rfig{d3dsim2}(a) shows time histories of simulations initialized with 
the $\rho_{q2} = 0.82$ case in \rfig{simbase}(a).
Total pressure $P$ 
is shown,
where $P$ is in arbitrary units. 
The resistive wall and ideal wall cases are included, as well as
resistive wall cases with the I-coil feedback models of \rfig{coils},\rfig{d3dwall}.
The perturbed $\psi$ with resistive wall and no feedback is shown in
\rfig{d3dfbk}(a), and the pressure in \rfig{d3dfbk}(b) having undergone a thermal quench.
The case with the worst feedback, with  $\psi_{f2}$,  also has a thermal quench. 
It evidently stabilizes the mode of \rfig{d3dfbk}(a),(b) and destabilizes the mode of
\rfig{d3dfbk}(c),(d).
 \rfig{d3dfbk}(c) 
shows that the mode connects with the upper part of the upper coil $\Psi_{UU}$  and
the lower part of the lower coil $\psi_{LL}$. This causes the mode to connect more strongly
with the resistive wall, 
The other two cases,  $\psi_{f1}$ and $\psi_{f1}+\psi_{f2},$  have very similar time evolution 
to an ideal wall. \rfig{d3dfbk}(e) shows perturbed $\psi$ using $\psi_{f1}$,
which indicates that the mode is pushed inward from the wall. \rfig{d3dfbk}(f)  
shows  pressure contours with significant  perturbations only near the plasma center. 

\rfig{d3dsim2}(b) shows time histories of simulations initialized with
the  $\rho_{q2} = 0.90$ case in \rfig{simbase}(a). The results are similar to
\rfig{d3dsim2}(a). A thermal quench occurs if $\psi_{fbk} = \psi_{f2},$ otherwise
$\psi_{f1}$ and $\psi_{f1}+\psi_{f2}$   are similar to an ideal wall.

With ideal wall or feedback which includes $\psi_{f1},$  the RWTM amplitude
remains small. 

 
\section{\Rdd{  MST feedback simulations }} \label{sec:mst}
\begin{figure}[h] 
\hspace{.5cm}
\begin{center}
\includegraphics[height=4.5cm] {\figdir/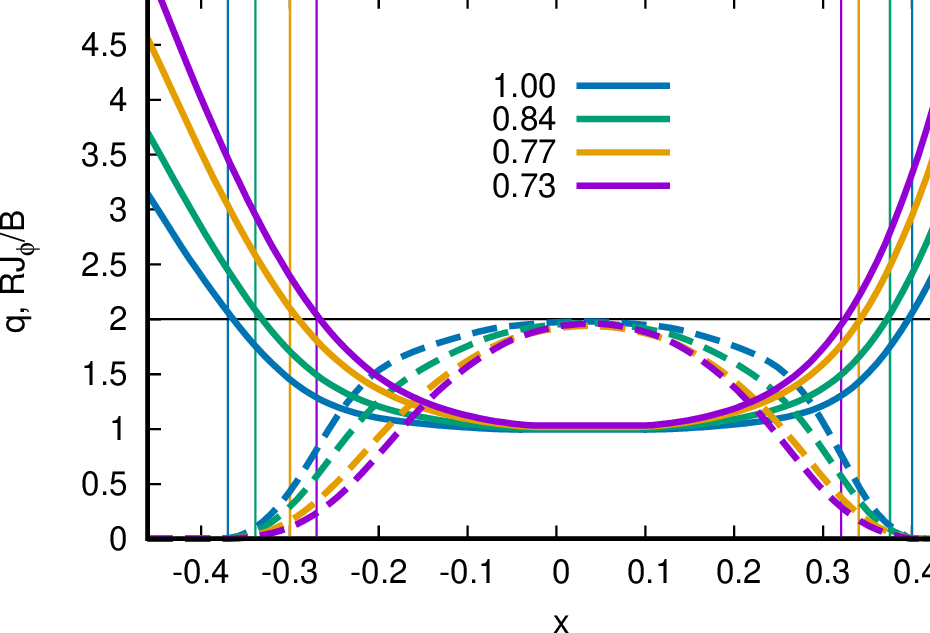}(a)
\includegraphics[height=4.5cm] {\figdir/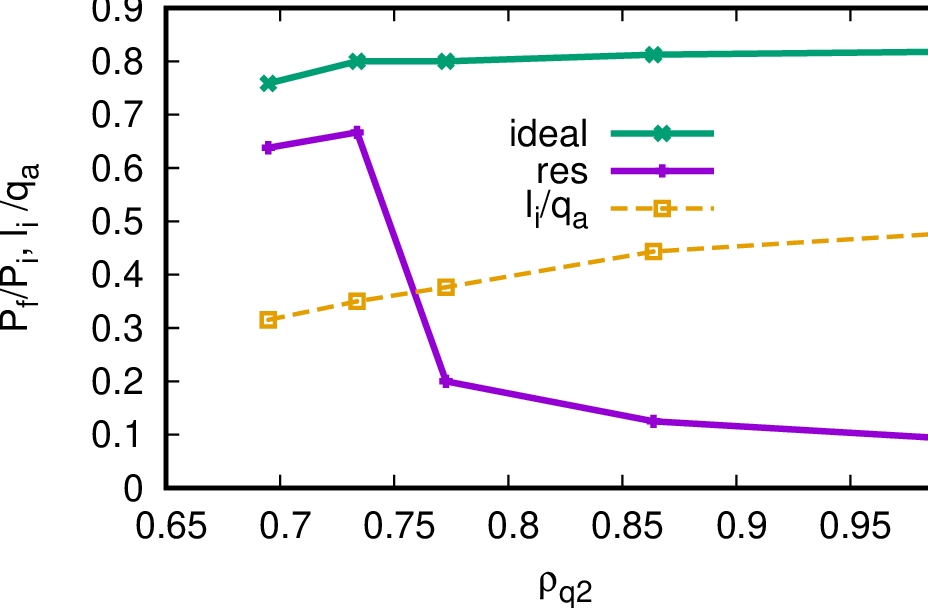}(b)
\end{center} \caption{ \it (a) $q(x)$ and $RJ_\phi(x)$ for  sequence of MST  equilibria with 
$0.7 \le \rho_{q2} \le 1,$ where $x = R - R_0.$ 
The curves are labelled with the values of $\rho_{q2}.$
Figure modified from \cite{nstx25}.
(b) Total pressure drop  $P_{final}/P_{initial}$ for ideal and resistive wall, $\rho_{q2}$
and $l_i / q_a$ 
as functions of $\rho_{q2}.$ Major disruptions occur for $\rho_{q2} \ge 0.75,$
$l_i / q_a > 0.35.$
}
\label{fig:mst}
\end{figure}
MST 
does not have disruptions because the wall is ideal on the experimental
  shot timescale. 
In  M3D
\cite{m3d} 
simulations of a sequence of equilibria \cite{nf-iaea24,model}  with $2.0 \le q_a \le 3.9,$ 
and $0.69 \le \rho_{q2} \le 1,$ the
resistive wall time was  made artificially short. 
The wall distance was increased to $\rho_w  = 1.2,$ like DIII-D.
Profiles of $q(x),$ $RJ_\phi(x)$ are shown in \rfig{mst}(a), where $x = R - R_0.$
The plasma extends to $x = 0.4,$ and the wall is at $x = 0.48.$ 
The curves are labelled with the values of $\rho_{q2}.$
As in \rfig{d3dsim}(a), the current density is small for $\rho > \rho_{q2}.$
Time history simulations \cite{nstx25,model} are summarized in \rfig{mst}(b), 
similar to \rfig{simbase}(b), 
giving the  total pressure drop  $P_{f}/P_{i}$ for ideal and resistive wall, 
and $l_i / q_a$,
as functions of $\rho_{q2}.$ 
Here $P_{f}$ is the total pressure after MHD activity, $P_{i}$ is the initial
total pressure. 
With a resistive wall, major disruptions occur  for $\rho_{q2} \ge 0.75.$
This is in agreement with the DIII-D database of Section 2. 
With an ideal wall, only minor disruptions occur. 

If feedback is applied to the resistive wall, an ideal wall can be
emulated.
\begin{figure}[h]
\begin{center}
\includegraphics[height=3.25cm]{\figdir/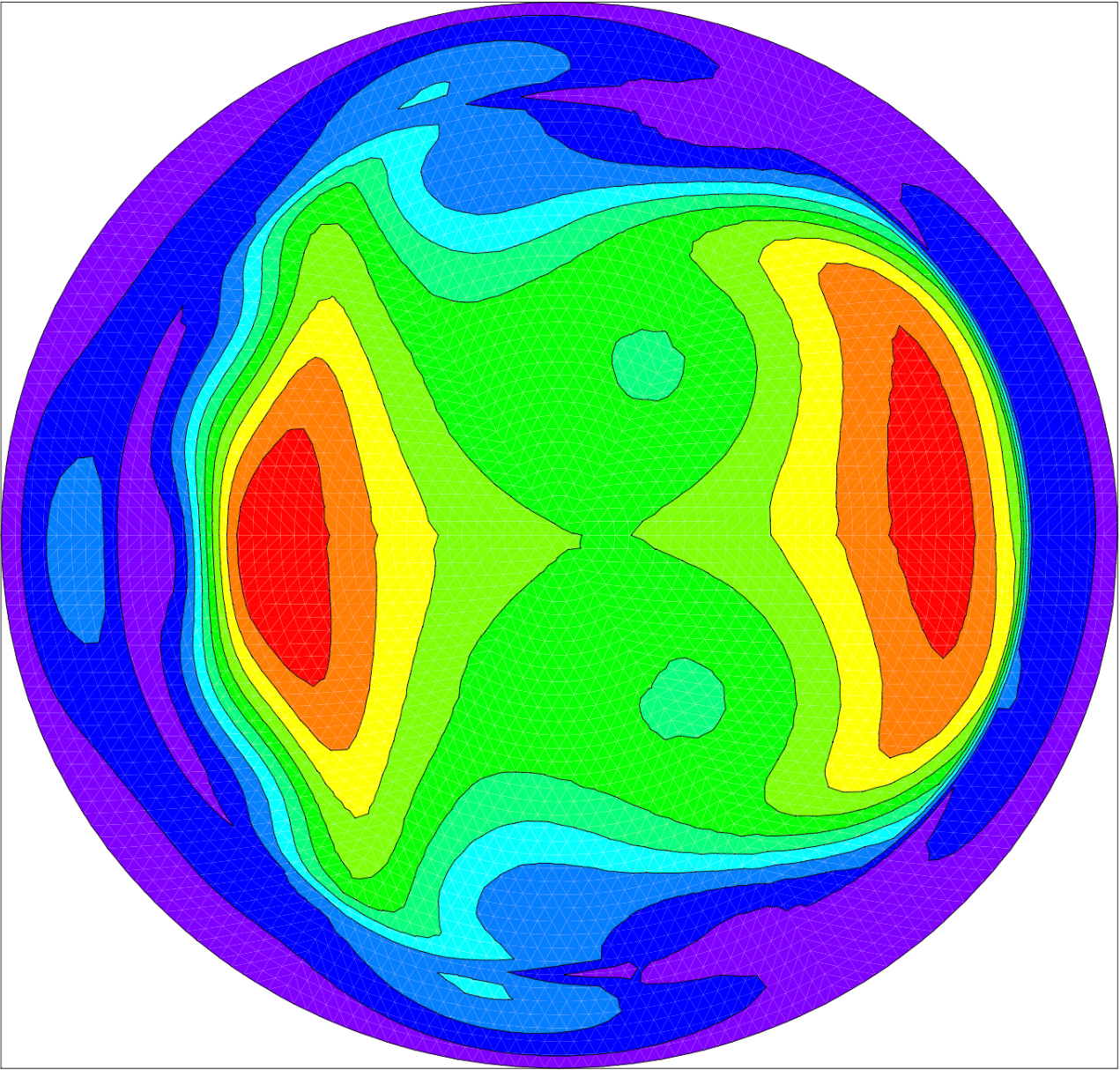}(a)
\includegraphics[height=3.0cm]{\figdir/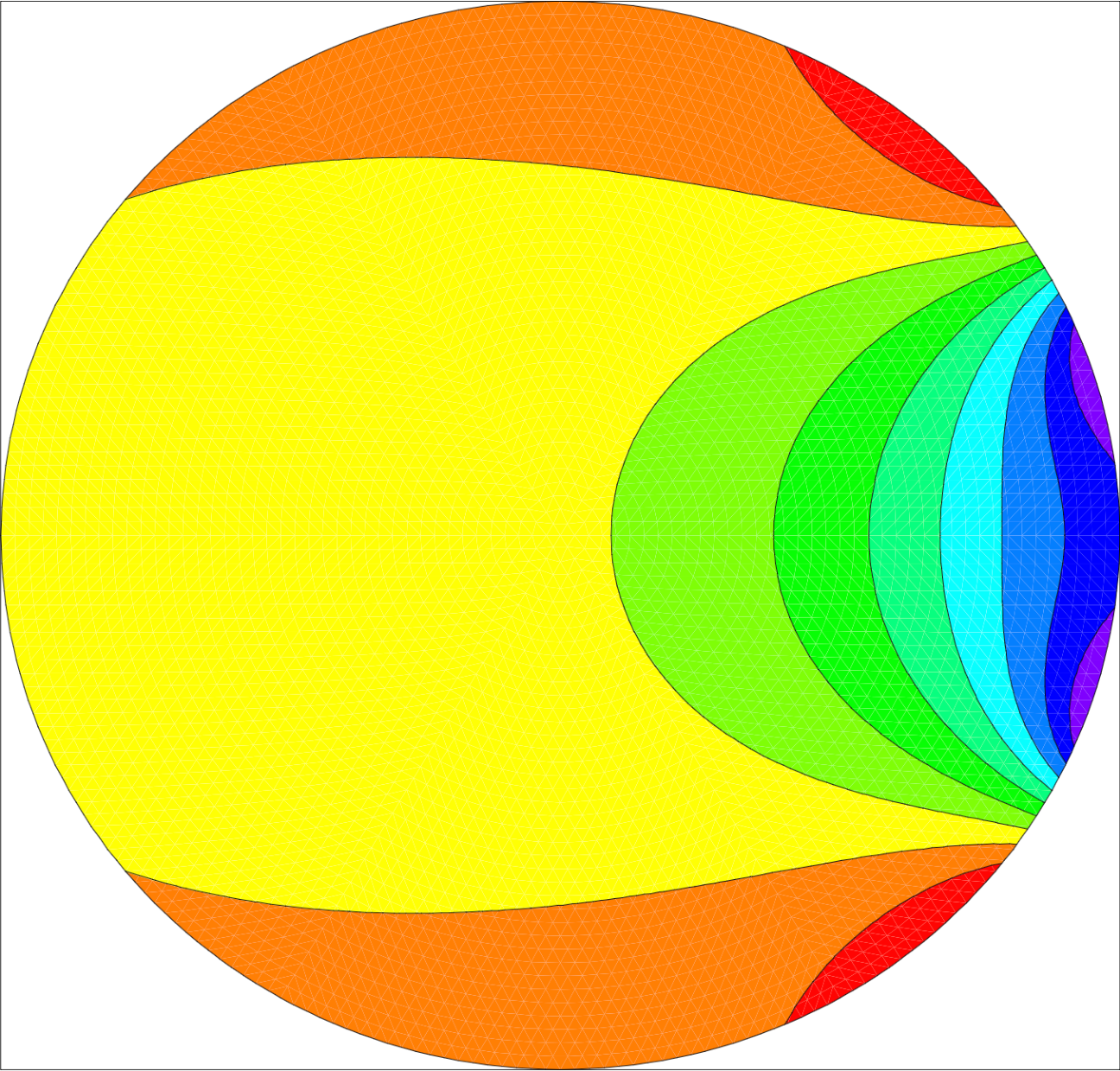}(b)
\includegraphics[height=3.0cm]{\figdir/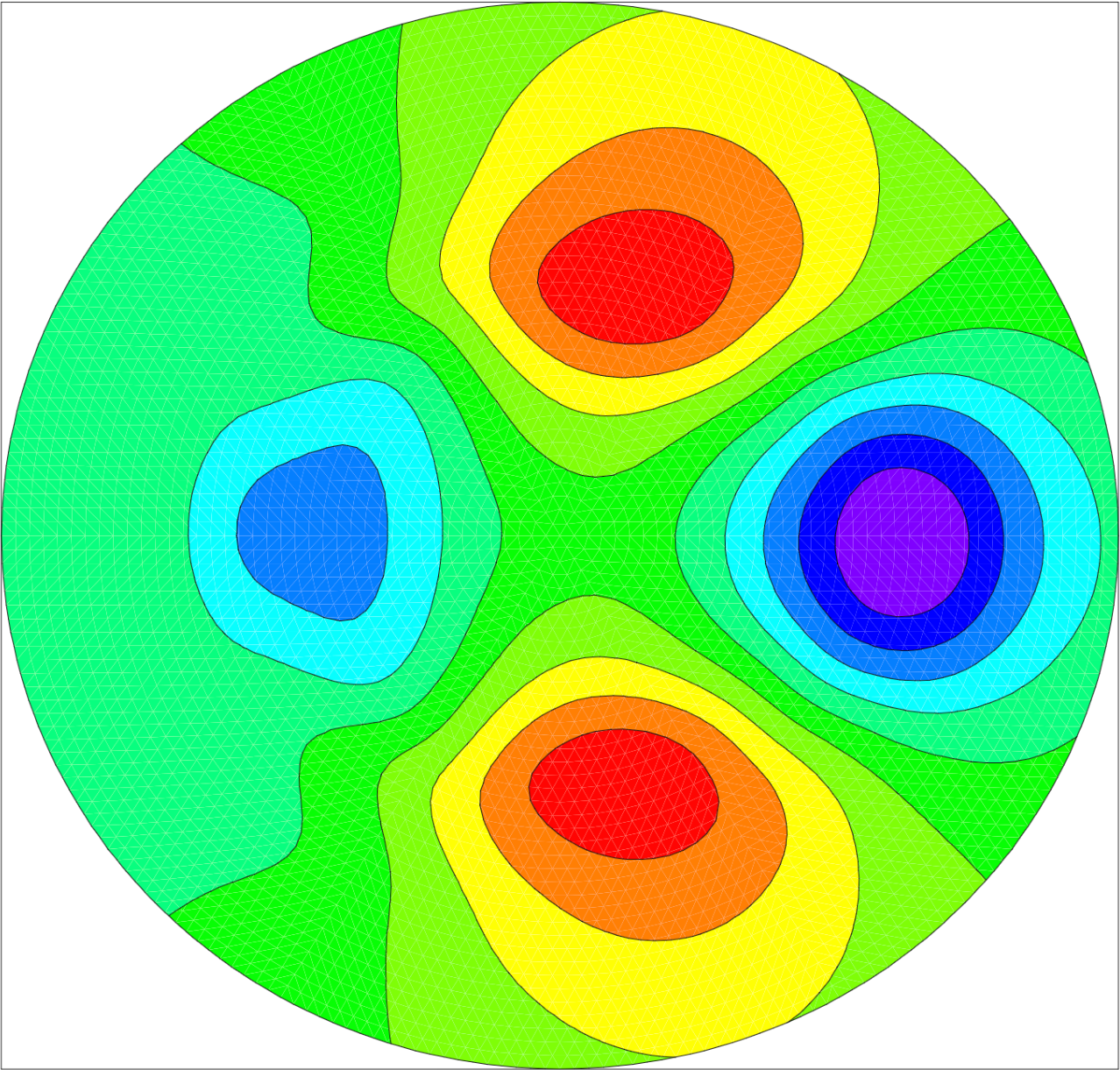}(c)
\includegraphics[height=3.0cm]{\figdir/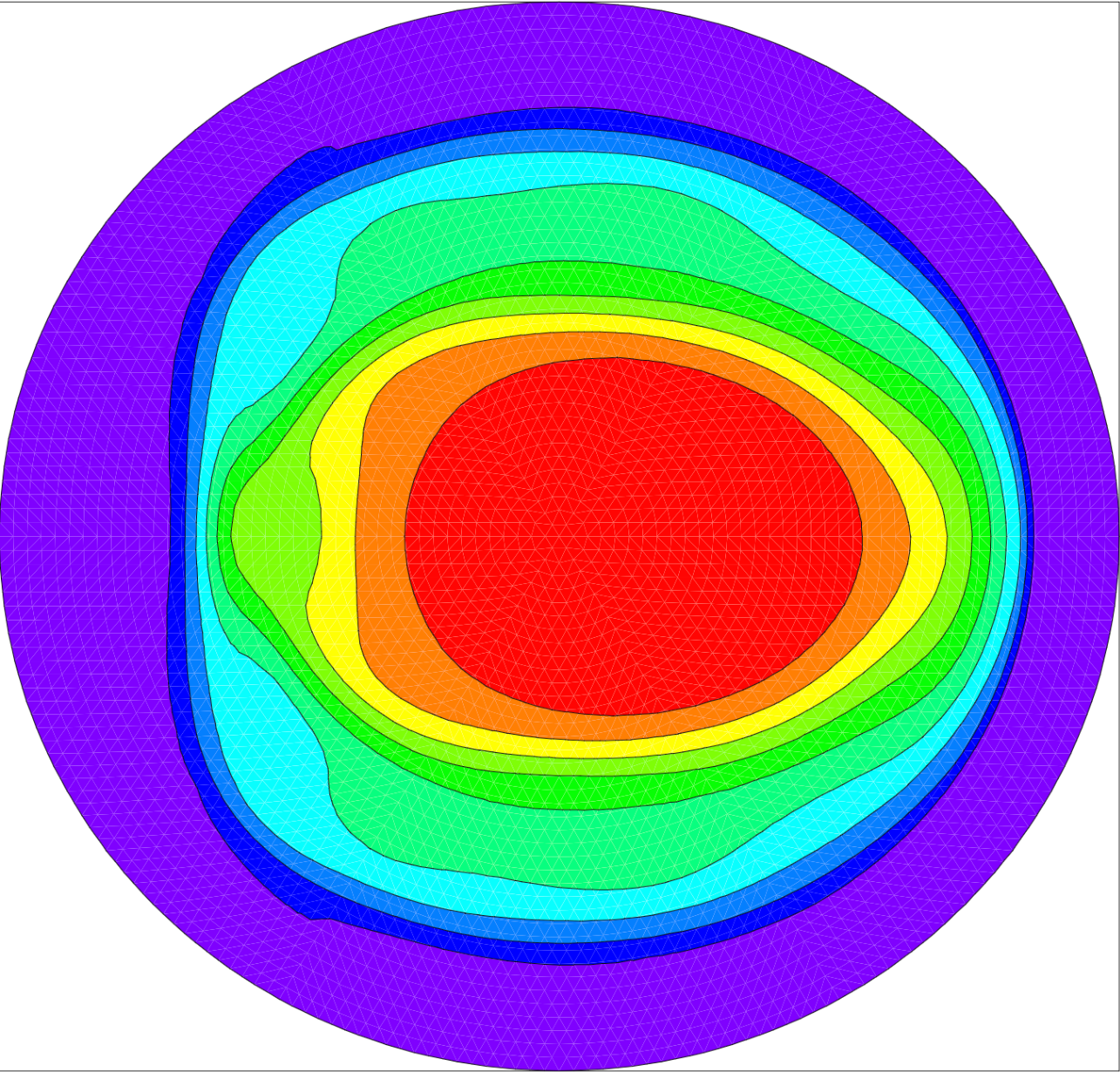}(d)
\end{center}
\caption{\it
(a) pressure contours with $\rho_{q2} = 0.77$, resistive wall,  without feedback, in a 
major disruption, reproduced from \cite{model} with AIP permission; 
(b) perturbed  magnetic flux $\psi_{fbk}$ from coils, 
similar to \rfig{coils}(c); compare with \rfig{lobe} (b);
(c) perturbed  magnetic flux $\psi$ with feedback pushing the mode away from the edge;
(d) pressure contours with feedback stabilization, limiting the
mode to much smaller amplitude.}
\label{fig:mstfeed}
\end{figure}
\rfig{mstfeed} (a) shows  pressure contours in a major disruption \cite{nstx25}, for 
$\rho_{q2} = 0.77$ and  resistive wall.
With no  feedback,
the  pressure contours show  a very large $(m,n) = (2,1)$ perturbation. 
The accompanying 
 $\psi$ perturbations are  shown in \rfig{lobe} (b).
Feedback is applied using 
magnetic flux $\psi_{f1}$ from  coils similar to DIII-D I-coils in \rfig{coils}(c),  shown in 
\rfig{mstfeed} (b). 
Comparing  with \rfig{lobe} (b) indicates that the feedback flux $\psi$  roughly
matches $\psi$ of  the perturbation at the wall.
\rfig{mstfeed} (c) shows  perturbed $\psi$  contours with feedback stabilization using the coils,
in which $\psi$ perturbations are pushed inward from the plasma edge.
\rfig{mstfeed} (d) shows  pressure contours with feedback stabilization using the coils, limiting the
mode to much smaller amplitude. 
The use of feedback prevents a major disruption.
 
 \section{\Rdd{\bf   Internal inductance criterion} } \label{sec:current} 
The $\rho_{q2}$  disruption criterion has a rather precise
interpretation as the wavelength from the mode rational surface to the wall.
The $l_i / q_{95}$ criterion  in the DIII-D disruption database
 is  less precise.
It is correlated  with the $\rho_{q2}$ criterion \req{fit}.
The relation appears to be a feature of equilibria
with $q_0 \approx 1,$ and low edge current at $\rho_{q2}$, as in \rfig{d3dsim}(a),   
\rfig{mst}(a).
These features can develop during disruption precursor evolution. 

\begin{figure}[h]
\hspace{.75cm}
\begin{center}
\includegraphics[height=4.2cm]{\figdir/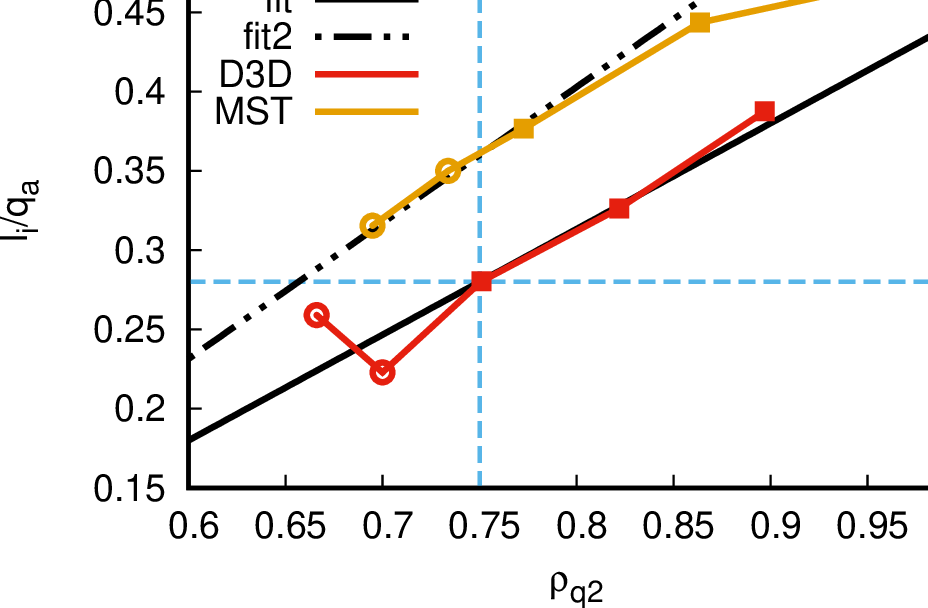}
\end{center}
\caption{\it
 nonlinear stability on $\rho_{q2}, l_i / q_{95}$ diagram. ``D3D" indicates DIII-D simulations
 from \rfig{simbase}(b). ``MST" is simulation data from \rfig{mst}(b). 
Filled square points have major
disruptions. Open circles do not. The line ``fit"  \req{fit} is a fit to the DIII-D
data and the simulations. The line ``fit2" is \req{fit} multiplied by $9/7.$ 
}
\label{fig:liq95}
\end{figure}

\rfig{liq95} 
presents simulation results from \rfig{simbase}(b) for DIII-D and \rfig{mst}(b) for MST.
The line labeled ``fit" is \req{fit}, which fits the DIII-D data of \rfig{sweeney}
along with simulations labelled ``D3D". 
Filled square points have major
disruptions. Open circles do not.
The simulations and data agree well.
The disruption criteria in the simulations agree with the experimental  data at the onset
condition $\rho_{q2} = 0.75,$ $l_i / q_{95} = 0.28.$
The MST simulations, presented in \rfig{mst}(b), 
agree  well with ``fit2", which is
the line ``fit" multiplied by a constant, $9/7.$
The alignment of the DIII-D $l_i / q_{95} $ curve with the ``fit" implies
that the profiles in \rfig{d3dsim}(a) are a reasonable approximation to the typical experimental
equilibria.
The  alignment of the MST simulations with ``fit2" indicates that they are related to
the DIII-D simulations and data.
The scaling factor between ``fit" and ``fit2" may be  due to  elongation, which is
 $\kappa = 1$ in MST, and $\kappa \approx 1.76$ in  DIII-D.
The ratio of the curves is $\kappa^{2/5}.$

During locked mode disruption precursors the plasma can develop
low temperature in the edge. 
This suppresses  the edge  current.
\rfig{d3dsim}(a), \rfig{mst}(a)  show that the current density is small in the edge region
$q \ge 2,$ while  the current density gradient at $\rho_{q2}$ is
substantial.
This has been  called a
``deficient edge" \cite{schuller} or
``minor disruption" \cite{wesson}. 
Edge cooling in DIII-D shot 154576  was produced by 
minor edge  disruptions  \cite{sweeney}.
Resistive ballooning turbulence can cause edge cooling, which might cause  
a density limit
\cite{ricci}. 

 \section {\Rdd{\bf Summary  }} \label{sec:summary} 
This paper deals with the stabilization of resistive wall tearing modes by
feedback.  
If the wall surrounding the plasma is resistive, RWTMs can cause a major disruption.
If the wall surrounding the plasma is ideal, 
only minor disruptions occur.
This is demonstrated for model DIII-D and MST equilibria,  derived from
experimental data.
It is shown that feedback can emulate an ideal wall.

This paper extends previous work 
\cite{nstx25,nf-iaea24,model} in
 several ways. Simulations representing the DIII-D locked mode disruption database
\cite{sweeney2017} are performed, with ideal and resistive wall boundary 
conditions.
With a resistive wall, major
disruptions occur when the $\rho_{q2}, l_i /q_{95}$ criteria are  satisfied.  
Similar results were obtained previously for a sequence of model MST equilibria. 
\cite{nstx25,model}.

Feedback
stabilization is studied with a more realistic model than previously \cite{nstx25,model}, 
based on DIII-D I-coils.
It is possible to prevent major disruptions  with feedback.
It is also shown that a certain choice of feedback is destabilizing, rather than
stabilizing, while another simple choice involving vertically symmetric feedback gives
very good results. A criterion using correlation of the feedback with the
mode allows an estimate of the required gain.

 It is generally acknowledged that disruptions are produced by
tearing modes, but the effect of a resistive wall has been less appreciated.
If the mode rational surface $\rho_{q2}$ is close enough to the wall to
interact with it, the tearing modes produce large perturbations near the
plasma edge. This causes enhanced parallel thermal transport and loss of
thermal energy across the plasma edge. The criterion was shown to be
$\rho_{q2} > 0.75$ in DIII-D and a model of MST.
If the wall surrounding the plasma is ideal, only minor disruptions occur.

Simulations with ideal wall boundary conditions have obtained disruptions, but
only for  highly unstable initial states. These would not occur in experiment,
where the unstable conditions develop in hundreds of milliseconds, allowing
time to relax the destabilizing drive.
An exception is mitigation by massive gas or shattered pellet injection, which
strongly perturbs the plasma.


With an ideal wall, minor, internal  disruptions can occur, caused by $(1,1)$ internal
kink or tearing modes. If $q_0 = 1,$ the $q = 1$ rational radius 
increases as $\rho_{q2}$ increases. 
This makes the internal
modes increase in   amplitude as $\rho_{q2}$ increases.  
The minor disruptions   
could be prevented by maintaining $q_0 > 1.$ 


The onset conditions $\rho_{q2},li/q_{95}$ are related to the current profile.
In the cases described here, unstable equilibria have $q_0 = 1$ and low edge
current. Experimentally,  low edge current is produced by edge cooling. This would tend
to suppress neoclassical tearing modes (NTM) \cite{lahaye}, which are driven by edge bootstrap
current.
The $\rho_{q2}$ criteria are the same in DIII-D and in the MST model. The
 $l_i/q_{95}$  criteria are
proportional, and apparently  depend on  plasma elongation. 

Feedback stabilization is preferable to the presently preferred method for
disruption mitigation, which is to inject high Z impurities in order to radiate
energy.  This method itself causes a disruption. A serious side effect is
excitation of runaway electrons, which may seriously damage the wall
surrounding the plasma. It seems sensible to avoid  this mitigation.

In future work, it is planned to improve the realism of the feedback model, and 
examine the feedback stabilization of RWTMs and RWMs for $\rho_{q2} \approx 1.$ 
It is also planned to examine the use of
 ITER resonant magnetic perturbation (RMP) coils for feedback.

{\bf Acknowledgement} Supported in part by USDOE grant DE-SC0020127. 

\small

\end{document}